\documentclass{article}

\usepackage{arxiv}
\usepackage{amsmath, amsfonts, amssymb}
\usepackage{times}
\usepackage{slashed}
\usepackage{tabularx}
\usepackage{multirow}
\usepackage[utf8]{inputenc} 
\usepackage[T1]{fontenc}    
\usepackage{hyperref}       
\usepackage{url}            
\usepackage{booktabs}       
\usepackage{amsfonts}       
\usepackage{nicefrac}       
\usepackage{microtype}      
\usepackage{graphicx}
\usepackage{siunitx}
\usepackage{braket}
\usepackage{bigints}
\usepackage{cite}
\usepackage{stfloats}
\usepackage{xcolor}
\definecolor{frenchblue}{rgb}{0.0, 0.45, 0.73}
\definecolor{RoyalRed}{RGB}{157, 16, 45}

\hypersetup{
  linkcolor  = frenchblue,
  urlcolor    = frenchblue,
  citecolor  = frenchblue,
  colorlinks = true,
  pdftitle     = {Electric dipole moments of baryons with bottom quarks},
  pdfauthor = {Y.~ \"Unal, Ulf-G. ~Mei{\ss}ner },
  pdfkeywords = {},
}

\makeatletter
\renewcommand{\maketitle}{\bgroup\setlength{\parindent}{0pt}
\begin{flushleft}
\textbf{\@title}
\@author
\end{flushleft}\egroup
}
\makeatother

\title{\Large{Electric dipole moments of baryons with bottom quarks} \\ }

\author{ \vspace*{10pt}
\textbf{Y.~\"Unal}$^{1,2,a}$, 	\textbf{D. ~Severt}$^{1,b}$,	\textbf{J.~de Vries}$^{3,4,c}$,
\textbf{C.~Hanhart}$^{5,d}$,	\textbf{Ulf-G.~Mei{\ss}ner}$^{1, 5,6,e}$\ \\ 
$^{1}$ \textit{Helmholtz-Institut f\"ur Strahlen- und Kernphysik and Bethe Center for
  Theoretical Physics Universit\"at Bonn, D-53115 Bonn, Germany} \\
$^{2}$\textit{Physics Department, \c{C}anakkale  Onsekiz Mart University 
17100 \c{C}anakkale, Turkey} \\
$^{3}$ \textit{Institute for Theoretical Physics Amsterdam and Delta Institute for
  Theoretical Physics, University of Amsterdam, Science Park 904, 1098 XH Amsterdam,
  The Netherlands} \\
  $^{4}$ \textit{Nikhef, Theory Group, Science Park 105, 1098 XG, Amsterdam, The Netherlands} \\
$^{5}$\textit{Institute for Advanced Simulation, Institut f\"ur Kernphysik and 
  J\"ulich Center for Hadron Physics, Forschungszentrum J\"ulich, D-52425 J\"ulich, Germany} \\
$^{6}$\textit{Tbilisi State University, 0186 Tbilisi, Georgia} \\
$^{a}$\href{mailto:yaseminunal@comu.edu.tr} {yaseminunal@comu.edu.tr},
$^{b}$\href{mailto:severt@hiskp.uni-bonn.de}{severt@hiskp.uni-bonn.de},
$^{c}$\href{mailto:j.devries4@uva.nl}{j.devries4@uva.nl},
$^{d}$\href{mailto:c.hanhart@fz-juelich.de}{c.hanhart@fz-juelich.de},
$^{e}$\href{mailto:meissner@hiskp.uni-bonn.de}{meissner@hiskp.uni-bonn.de	}	
}

\begin{document}

\clearpage
\maketitle
\thispagestyle{empty}

\textbf{\textit{Abstract}}
Triggered by experimental prospects to measure electromagnetic dipole moments of baryons containing a
bottom quark, we calculate the CP-odd electric dipole moments (EDMs) of spin-$1/2$ single-bottom baryons.
We consider CP-violating dimension-six operators in the Standard Model Effective Field Theory that
involve bottom quarks, and apply heavy-baryon chiral perturbation theory to compute the EDMs of
several baryons. We discuss the expected size of the EDMs for beyond-the-Standard Model physics
appearing at the TeV scale. 


\section{Introduction}

Experiments aiming to detect permanent
electric dipole moments (EDMs) set strong bounds
on flavor-diagonal mechanisms that simultaneously violate time-reversal (T) and parity (P)
(and thus CP symmetry 
if we take CPT to be a good symmetry of nature).
For instance, the strongest constraints on the QCD $\bar \theta$-term arise from measurements of 
the EDMs of the neutron and the ${}^{199}$Hg atom \cite{nEDM:2020crw,Griffith:2009zz}. 
In addition, EDM experiments strongly constrain possible sources of CP violation
from beyond-the-Standard-Model (BSM) physics. While EDMs have been calculated in a plethora of different
BSM models,  BSM CP violation can be described more systematically in the framework of the Standard Model
Effective Field Theory 
(SMEFT)~\cite{Grzadkowski:2010es} under the reasonable assumption that the scale of BSM physics,
$\Lambda$, lies well beyond the electroweak scale, $v \simeq 250\,$GeV.

A lot of effort has gone into more and more accurate calculations of EDMs of systems containing
first-generation valence quarks 
such as nucleons, nuclei, atoms, and molecules
\cite{ deVries:2010ah,deVries:2012ab, Bsaisou:2014oka, Yamanaka:2020kjo,Kley:2021yhn}. The associated
experiments are mainly sensitive to CP-odd SMEFT operators
containing light quarks (and leptons, but we will not pursue leptonic CP violation in this work).
For instance, the non-observation of a neutron EDM
sets stringent limits on the electric and chromo-electric dipole moments of up and down quarks and various
four-quark interactions~\cite{Dekens:2014jka}. 
The experimental limits are so stringent, that the same experiments also indirectly constrain CP
violation in interactions involving heavier quarks.
For instance, a chromo-electric dipole moment of a bottom or top quark, induced at the scale $\Lambda$
in some BSM theory, will in turn induce
chromo-electric dipole moments of light quarks and gluons due to renormalization-group evolution to lower
energies and threshold effects when the
heavier quarks are integrated out. Systematic studies of the resulting indirect limits have appeared in several places in the literature see e.g. \cite{Braaten:1990zt,Chien:2015xha,Gisbert:2019ftm,Haisch:2021hcg}.

Although those indirect limits are already valuable, more direct information on CP-violating
interactions involving heavy quarks would be welcome. 
First of all, additional observables would help in setting global constraints leaving less room for
possible cancellations among various sources. Second,  as soon as a non-zero
EDM will be found, hopefully in the near future, additional information is needed to pin down the underlying
source of CP violation. Third, while operators with heavy quarks contribute to first-generation EDMs, the
contributions are loop suppressed and sometimes involve small dimensionless numbers such as Cabibbo-Kobayashi-Maskawa (CKM) matrix elements
or light-quark Yukawa couplings. Finally, and arguably most importantly, plans are being discussed to
measure EDMs of baryons with a heavy valence quarks directly. For instance,
Refs.~\cite{Fomin:2017ltw,Aiola:2020yam,Bagli:2017foe} discuss the prospects of measuring EDMs of
charm and bottom baryons. Further discussions on the mechanism of CP violation resulting from the
QCD $\theta$-term in the charm baryon sector can be found in \cite{Unal:2020ezc}. In this work, we
calculate the EDMs of spin-$1/2$ bottom baryons in the framework of the SMEFT. In this way, we
can determine what is the sensitivity of potential future measurements on the scale of BSM physics,
and whether different baryons have a different sensitivity to various CP-violating SMEFT operators.

This paper is organized as follows. In Sect.~\ref{dim6b} we discuss dimension-six SMEFT CP-violating
operators involving bottom quarks. In Sect.~\ref{secChPT} we discuss how to match these operators to the
hadronic level using chiral perturbation theory focusing on the operators most relevant for our EDM
calculations. In Sect.~\ref{sec:loop} we perform the calculation of the EDMs of bottom-quark baryons
at leading order for each source of CP violation. We discuss the expected sizes of EDMs in
Sect.~\ref{sec:size} and conclude in Sect.~\ref{sec:con}. Several appendices are devoted to technical issues.

\section{CP-violating operators involving bottom quarks}\label{dim6b}

We start with listing CP-violating operators involving $b$ quarks at the quark level. We focus on
operators with at least
one $\bar b \Gamma b$ bilinear, where $\Gamma$ is a general Lorentz-structure, while the remaining
fields are light quarks or gauge or scalar bosons. 
Operators with more $b$ quark fields lead to
suppressed EDMs of systems containing a single $b$ valence quark in the
same way as $b$ quark effects are suppressed in light states. We do not consider operators with just light quarks
even though they would contribute to b-quark containing baryons. The reason being that the limits on these CP-odd operators
from traditional EDM experiments, such as those for the neutron EDM, are very stringent. 

At low energies, right above the $b$-quark threshold, the effective P- and T-violating
dimension-six operators of relevance here reads
~\cite{Grzadkowski:2010es, deVries:2012ab, Bsaisou:2014goa} 
\begin{equation}\
\begin{aligned}
\mathcal{L}_{b,  \mathrm{qEDM}}^{(6)}=~& d_b \bar{b}~ \sigma^{\mu \nu} \gamma_5 b ~F_{\mu \nu}\,, \\
\mathcal{L}_{b,  \mathrm{qCEDM}}^{(6)}=~& \tilde{d}_b \bar{b}~\sigma^{\mu \nu} \gamma_5 \lambda^a b~G_{\mu \nu}^a\,, \\
\mathcal{L}_{b,  \mathrm{4q}}^{(6)}=~& i \mu_1^{ub} (\bar{u}u\bar{b}\gamma_5 b + \bar{u}\gamma_5u\bar{b}b -
\bar{b} \gamma_5 u \bar{u}b - \bar{b}u \bar{u} \gamma_5 b) + i \mu_1^{db} (\bar{d}d\bar{b}\gamma_5 b
+ \bar{d}\gamma_5d\bar{b}b \\
&- \bar{b} \gamma_5 d \bar{d}b - \bar{b}d \bar{d} \gamma_5 b) + i \mu_1^{sb} (\bar{s}s\bar{b}
\gamma_5 b + \bar{s}\gamma_5s\bar{b}b - \bar{b} \gamma_5 s \bar{s}b - \bar{b}s \bar{s} \gamma_5 b) \\
                                                    &+i \mu_8^{ub} (\bar{u} \lambda^a u\bar{b}\gamma_5 \lambda^a b + \bar{u}\gamma_5 \lambda^a u\bar{b} \lambda^a b - \bar{b} \gamma_5 \lambda^a u \bar{u} \lambda^a b - \bar{b} \lambda^a u \bar{u} \gamma_5 \lambda^a b) \\
                                                    &+i \mu_8^{db} (\bar{d} \lambda^a d\bar{b}\gamma_5 \lambda^a b + \bar{d}\gamma_5 \lambda^a d\bar{b} \lambda^a b - \bar{b} \gamma_5 \lambda^a d \bar{d} \lambda^a b - \bar{b} \lambda^a d \bar{d} \gamma_5 \lambda^a b) \\
                                                    &+i \mu_8^{sb} (\bar{s} \lambda^a s\bar{b}\gamma_5 \lambda^a b + \bar{s}\gamma_5 \lambda^a s\bar{b} \lambda^a b - \bar{b} \gamma_5 \lambda^a s \bar{s} \lambda^a b - \bar{b} \lambda^a s \bar{s} \gamma_5 \lambda^a b)\,,\\
	\mathcal{L}_{b,  \mathrm{4qLR}}^{(6)}=~& i \nu_{1}^{ub} V_{ub} (\bar{b}_L\gamma_{\mu} u_L \bar{u}_R\gamma^{\mu} b_R)-i\nu_{1}^{ub} V_{ub}^* (\bar{b}_R\gamma_{\mu} u_R \bar{u}_L\gamma^{\mu} b_L) \\
                                                   &+i \nu_{8}^{ub} V_{ub} (\bar{b}_L \gamma_{\mu} \lambda^a u_L \bar{u}_R \gamma^{\mu} \lambda^a b_R)-i\nu_{8}^{ub} V_{ub}^* (\bar{b}_R\gamma_{\mu} \lambda^ a u_R \bar{u}_L\gamma^{\mu} \lambda^ a b_L)\,,
	\label{pt_vio_op}
	\end{aligned}
\end{equation}
where $V_{ub}$ is an element of the CKM matrix, $F_{\mu \nu}$
and $G_{\mu \nu}^a$ are the electromagnetic and the gluon field-strength tensors, respectively. 

The bottom-quark EDM (qEDM) and bottom-quark chromo-EDM (qCEDM) operators arise from the following dimension-six operators in the SMEFT Lagrangian 
\begin{eqnarray}
\mathcal L_{\mathrm{4q}}=C^{bB}\,(\bar Q_3 \sigma^{\mu\nu} b_{R_b}) H B_{\mu\nu} + C^{bW}\,(\bar Q_3 \sigma^{\mu\nu} \tau^a b_{R_b}) H W^a_{\mu\nu}+ C^{bG}\,(\bar Q_3 \sigma^{\mu\nu} \lambda^a b_{R_b}) H G^a_{\mu\nu} + {\rm h.c.}
\label{eq:EFT4fermion}
\end{eqnarray}
where $Q_3$ denotes a left-doublet of third-generation quarks, $H$ is the Higgs doublet, and
$B_{\mu\nu}$ and $W_{\mu\nu}^a$ denote, respectively, the $U(1)_Y$ and $SU(2)_L$ field strengths.
To preserve gauge invariance, the SMEFT dipole operators involve a Higgs field in the SMEFT Lagrangian.
Eq.~\eqref{pt_vio_op} is subsequent to electroweak symmetry breaking where we have replaced the Higgs
field by its vacuum expectation value. The bottom qEDM arises from a linear combination of $U(1)_Y$
and $SU(2)_L$ dimension-six dipole operators (there is in principle an associated dipole operator
coupled to $Z$ and $W^\pm$ bosons that play no role in our analysis). In most models of BSM physics,
the dipoles scale with the bottom quark Yukawa and we expect $d_b, \tilde{d}_b \sim m_b/\Lambda^2$.
These dipole operators are generated in various classes of BSM physics ranging from
supersymmetric scenarios \cite{Nakai:2016atk}, to two-Higgs doublet models \cite{Jung:2013hka},
to leptoquarks \cite{Dekens:2018bci}.

The four-quark (4q) operators in $\mathcal{L}_{b,  \mathrm{4q}}^{(6)}$ are induced from gauge invariant
operators of the form  
\begin{eqnarray}
\mathcal L_{\mathrm{4q}}=C^{abcd}_{\mathrm{4q}}\,(\bar Q_a^I  u_{R_b}) \epsilon_{IJ}(\bar Q_c^J  d_{R_d}) +{\rm h.c.}+\dots\,,
\label{eq:EFT4fermion}
\end{eqnarray}
where the ellipses denote terms with additional color structure, and $abcd$ are generation indices.
These operators
induce $\mathcal{L}_{b,  \mathrm{4q}}^{(6)}$ for the generation indices $a=b=\{1,2\}$ and $c=d=3$ or $a=d=3$
and $b=c=\{1,2\}$ (the operator in Eq.~\eqref{pt_vio_op} are associated to the former generation configuration.
The second configuration leads to very similar low-energy operators and the analysis presented here
will be the same)
and additional operators involving top quarks that play no role at low energies. We expect
$\mu^{ub,db,sb}_{1,8}\sim 1/\Lambda^2$. For instance, the CP-odd four-quark operators are induced
in models of leptoquarks in which case $\Lambda$ is related to the mass of the exchange
leptoquark \cite{Dekens:2018bci}. 

The four-quark operators in $\mathcal{L}_{b,  4qLR}^{(6)}$ are induced from the gauge-invariant operator 
\begin{eqnarray}
\mathcal L_{\mathrm{4qLR}}=
C^{ab}_{\mathrm{4qLR}}\,\left(\tilde{H}^{\dagger} D_{\mu} H\right) \, \bar{u}^a_R \gamma^\mu  b^b_R+{\rm h.c.}~.
\end{eqnarray}
After electroweak symmetry breaking this operator leads to a right-handed charged current. This operator is usually called four-quark left-right (4qLR) operator. The interactions in
$\mathcal{L}_{b,  4qLR}^{(6)}$ are generated when the $W$ boson is integrated out at tree level between
quarks giving rise to the additional factor of $V_{ub}$. By power counting $\nu_{1,8}^{ub} \sim
v^2/(m_W^2 \Lambda^2) \sim 1/\Lambda^2$. An example where this operator is generated is the minimal
left-right symmetric model \cite{Dekens:2014jka}.

\section{Chiral perturbation theory for bottom baryons}\label{secChPT}

The way to include heavy bottom quarks into standard Chiral Perturbation Theory (ChPT) is known for some
time \cite{Cheng:1993kp, Yan:1992gz}. In the SU(3) flavor representation the spin-1/2 anti-symmetric
triplet and symmetric sextet bottom baryon states are denoted by the following matrices,  respectively, ~
\begin{equation}
  	B_{\bar{3}}=
  \begin{pmatrix}
               0                    & \Lambda_{b}^0        & \Xi_{b}^0 \\
    - \Lambda_{b}^0      &        0                       & \Xi_{b}^- \\
    - \Xi_{b}^0                & -\Xi_{b}^-                &     0
  \end{pmatrix},\quad  	
  	B_6=
  \begin{pmatrix}
    \Sigma_{b}^{+}                           & \frac{\Sigma_{b}^0}{\sqrt{2}}           & \frac{\Xi_{b}^{'0}}{\sqrt{2}}  \\
    \frac{\Sigma_{b}^{0}}{\sqrt{2}}    & \Sigma_{b}^-                                  & \frac{\Xi_{b}^{'-}}{\sqrt{2}}  \\
    \frac{\Xi_{b}^{'0}}{\sqrt{2}}           & \frac{\Xi_{b}^{'-}} {\sqrt{2}}           & \Omega_{b}^{-}
  \end{pmatrix}.
	\end{equation}
The Goldstone boson octet is given by 
\begin{equation}
	\phi=
	\begin{pmatrix}
	\frac{1}{\sqrt{2}}\pi^0+ \frac{1}{\sqrt{6}}\eta                      & \pi^+        & K^+ \\
	\pi^-      &        -\frac{1}{\sqrt{2}}\pi^0+ \frac{1}{\sqrt{6}}\eta                        & K^0 \\
	K^-                & \bar{K}^0                &     - \frac{2}{\sqrt{6}}\eta       
	\end{pmatrix}\,,
	\end{equation} 
and we define 
	\begin{equation}
	U=u^2=\text{exp}\left(i \frac{\phi}{F_{\pi}}\right) , 
	\end{equation} 
where $F_\pi$ is the pion decay constant. The relevant P- and T-conserving free and interaction
Lagrangians up to the second chiral order in a covariant formalism are given by
~\cite{Yan:1992gz,  Xiabng:2018qsd, Jiang:2014ena, Borasoy:2000pq} 
\begin{equation}
\begin{aligned}
  \mathcal{L}_{\text{free}}^{(1)}= ~& \frac{1}{2} \langle \bar{B}_{\bar{3}}(i \slashed{D}-m_{\bar{3}})
  B_{\bar{3}} \rangle
+\langle \bar{B}_{6}(i \slashed{D}-m_{6}) B_{6} \rangle\,,\\                                                                                                                
\mathcal{L}_{\text{int}}=~ ~ &\frac{g_1}{2} \langle \bar{B}_{6}\slashed{u} \gamma_5 {B}_{6} \rangle+\frac{g_2}{2} \langle \bar{B}_{6}\slashed{u} \gamma_5 {B}_{\bar{3}}+h.c. \rangle+\frac{g_3}{2} \langle \bar{B}_{\bar{3}}\slashed{u} \gamma_5 {B}_{\bar{3}} \rangle \,,  \\                                                                                                                                                    
\mathcal{L}_{B \gamma}^{(2)}= ~&w_1 \langle \bar{B}_{\bar{3}} \sigma^{\mu \nu}
F_{\mu \nu}^+B_{\bar{3}} \rangle
+w_2\langle \bar{B}_{6} \sigma^{\mu \nu} F_{\mu \nu}^+ B_{6} \rangle + w_3 \langle \bar{B}_{6} \sigma^{\mu \nu} F_{\mu \nu}^+B_{3}
+ h.c. \rangle+w_4 \langle \bar{B}_{\bar{3}} \sigma^{\mu \nu} B_{\bar{3}} \rangle \langle F_{\mu \nu}^+\rangle \\
&+w_5 \langle \bar{B}_{6} \sigma^{\mu \nu} B_{6} \rangle\langle F_{\mu \nu}^+\rangle \,.
\label{Meson-baryon Lagr}				    
\end{aligned}
\end{equation} 
Here, $D_{\mu}$ is the covariant derivative defined as
\begin{equation}
	D_{\mu}^{} B = \partial_{\mu}^{} B + \Gamma_{\mu}^{} B + B \Gamma_{\mu}^{T} , \quad \Gamma_{\mu} = \frac{1}{2} \left[ u^{\dagger} ( \partial_{\mu} - i r_{\mu} ) u + u ( \partial_{\mu} - i l_{\mu} ) u^{\dagger} \right] ,
\end{equation}
and $u_{\mu}$ is the standard chiral Vielbein 
\begin{equation}
	u_{\mu} = i \left[ u^{\dagger} ( \partial_{\mu} - i r_{\mu} ) u - u ( \partial_{\mu} - i l_{\mu} ) u^{\dagger} \right] , 
\end{equation}
where $r_{\mu}$ and $l_{\mu}$ denote external right- and left-handed sources. Also, we have  
\begin{equation}
	F_{\mu \nu}^+ = u^{\dagger} Q_B F_{\mu \nu} u + u Q_B F_{\mu \nu} u^{\dagger} , 
\end{equation}
with the bottom baryon charge matrix \cite{Guo:2008ns}
\begin{equation}
	Q_B = \frac{e}{2} \text{diag} ~ ( 1 ,-1,-1) .
\end{equation} 
The prefactors $g_{1-3}$ and $w_{1-5}$ are low-energy constants (LECs). $g_2$ is calculated using
the widths of the heavy baryons. $g_1$ and $g_3$ are related to $g_2$ with the help of the quark
model and heavy quark spin flavor symmetry~\cite{Yan:1992gz,Guo:2008ns,Jiang:2014ena, Jiang:2015xqa}.
Due to heavy quark spin symmetry, the vertex $B_{\bar{3}}B_{\bar{3}} \phi $ is forbidden and the term has
to vanish, i.e. $g_3=0$. This result can be deduced from angular momentum and parity conservation
arguments (see e.g.~\cite{Yan:1992gz}). The conventional magnetic moment couplings, $w_{1-5}$ are determined
from fits to calculations to baryon magnetic moments in \cite{Meissner:1997hn, Kubis:2000aa}.
However, in the present calculation, they do not contribute to the EDMs at the order we work.
The numerical values of the contributing couplings are given in Section~\ref{sec:size}.

\subsection{Construction of the effective CP-violating Lagrangian}

We now turn to the construction of the effective Lagrangian on the hadron level arising from the
dimension-six terms in Eq.~\eqref{pt_vio_op}. 
The first operator we want to look at is the bottom-quark EDM (qEDM) which does not contain any
light quarks but only the heavy $b$-quark. As it already contains the electromagnetic field strength
tensor $F_{\mu \nu}$, it directly induces EDMs of baryons containing bottom quarks. We find only
two terms in the leading chiral Lagrangian corresponding to EDMs of the anti-triplet and sextet of
bottom-quark baryons.

Next, we discuss the bottom-quark CEDM (qCEDM). Similarly to the qEDM, there is no light quark content
in the Lagrangian and, instead of $F_{\mu \nu}$, we have the gluon field strength tensor $G_{\mu \nu}^{a}$.
The fact that this term contains only heavy quarks and $G_{\mu \nu}^{a}$ makes this term (like the qEDM)
a chiral singlet, i.e. it is invariant under chiral SU(3) transformations. 
In standard ChPT, there is no fundamental building block that transforms as a chiral singlet. 
Therefore, we have to introduce a new fundamental block $\beta^+$, which gives the proper transformation behaviour, and a partner building block $\beta^-$, which transforms accordingly and explicitly violates P and T. This procedure works analogously to the definition of the building blocks $\chi_+$ and $\chi_-$ in ChPT. However, in contrast to the building blocks $\chi_{\pm}$, the chiral singlet $\beta^+$ cannot introduce any further structure containing Goldstone boson fields. In fact, it can be shown that $\beta^+$ can only enter the effective Lagrangian as an overall constant. In a similar fashion, one can also deduce that a P- and T-violating chiral singlet term $\beta^-$ will always vanish. There is simply no constant that can violate P and T.
Despite $\beta^+$ being a constant, we still have to treat it like a building block. To construct the effective Lagrangian on the hadron level, we need to combine $\beta^+$ with other ChPT building blocks that violate CP. This procedure leads to the terms given below. For more information see e.g. Refs.~\cite{Bsaisou:2014goa, Bsaisou:2014oka}.


The next contributions we investigate are the four quark interaction terms (4q-operators). These terms
need a little extra treatment, since they not just include the heavy bottom quark, but also the
light quarks $u$, $d$ and $s$. Due to the presence of the light quarks, we have to study how the 4q-terms
transform under chiral transformations. 
To obtain the transformation properties of $\mathcal{L}_{b,  \mathrm{4q}}^{(6)}$ under chiral SU(3)
transformations, we first express the non-mixing $\mu_1$ terms of the operator as follows
\begin{eqnarray}
i \mu_1^{ub} (\bar{u}u\bar{b}\gamma_5 b + \bar{u}\gamma_5u\bar{b}b ) + i \mu_1^{db} (\bar{d}d\bar{b}\gamma_5 b + \bar{d}\gamma_5d\bar{b}b )+ i \mu_1^{sb} (\bar{s}s\bar{b}\gamma_5 b + \bar{s}\gamma_5s\bar{b}b).  
\label{nomix}
\end{eqnarray}     
These terms have the structure
\begin{eqnarray}
i \bar{q} \mathcal{M}_1 q ~ (\bar {b} \gamma_5 b) + i \bar{q} \mathcal{M}_1 \gamma_5 q ~ (\bar{b}b),
\label{mass1}
\end{eqnarray}
in terms of the quark column vector $q = (u\,d\,s)^T$ and
\begin{equation}
\mathcal{M}_1=
\begin{pmatrix}
\mu_1^{ub} & 0 & 0 \\
0 & \mu_1^{db} & 0 \\
0 & 0 & \mu_1^{sb} 
\end{pmatrix}.
\label{matrix}
\end{equation}
For the light quarks, Eq. (\ref{mass1}) has the structure of a mass term
in ordinary ChPT, because the term containing the $b$ quarks is a SU(3) singlet and does not transform at
all. The $\mathcal{M}_1$ matrix will therefore act as a new scalar source, similar to the quark mass
matrix in standard ChPT, while the explicit insertions of the $b$-quark field allow for the appearance
of the heavy bottom baryon matrices $B_{\bar{3}}$ and $B_{6}$ in the effective Lagrangian. 

The mixing terms in the 4q Lagrangian, 
\begin{eqnarray}
- i \mu_1^{ub} ( \bar{b} \gamma_5 u \bar{u}b + \bar{b}u \bar{u} \gamma_5 b) - i \mu_1^{db} (\bar{b} \gamma_5 d \bar{d}b + \bar{b}d \bar{d} \gamma_5 b) - i \mu_1^{sb} ( \bar{b} \gamma_5 s \bar{s}b + \bar{b}s \bar{s} \gamma_5 b) ,
\label{mix}
\end{eqnarray}
can be treated in an analogous way. If we use the identities 
\begin{eqnarray}
\bar{q}q=&\bar{u}u+\bar{d}d+\bar{s}s,  \nonumber\\
q\bar{q}= & \begin{pmatrix}
u\bar{u} & u\bar{d} & u\bar{s} \\
d\bar{u} &d\bar{d} & d\bar{s} \\
s\bar{u} & s\bar{d} & s\bar{s}
\end{pmatrix},  \nonumber\\
\langle q\bar{q} \rangle=&u\bar{u}+d\bar{d}+s\bar{s}~, 
\label{scalar}
\end{eqnarray}
we can express Eq.(\ref{mix}) together with Eqs.(\ref{matrix}, \ref{scalar}) as 
\begin{eqnarray}
- i \bar{b} \gamma_5 \langle (\mathcal{M}_1q)\bar{q} \rangle b - i \bar{b} \langle
(\mathcal{M}_1q)\bar{q} \rangle \gamma_5 b .
\end{eqnarray}
Using the cyclic property of the trace one observes that these mixing terms transform again
like a mass term. Thus, we can use the same procedure like in the non-mixing case to
obtain the effective Lagrangian.  

For the 4q-operators, the $\mu_1$- and $\mu_8$-terms have identical chiral symmetry properties. 
While these terms are distinguishable on the quark-level, at low energies the resulting chiral Lagrangians
are identical. We are not able to distinguish them without nonperturbative information  about the
associated low-energy constants. 
The effective Lagrangian from the $4q$-operator will therefore combine the effects of the $\mu_1$
and $\mu_8$ terms. 

The last terms we have to discuss are the 4qLR-terms. Similarly to the 4q-operator one can
reproduce the transformation rules for the 4qLR-operator. First we take the $\nu_1^{ub}$-terms and
use Fierz identities to rewrite the left- and right-handed components of the quark fields. Then,
we arrange the resulting terms, like before,
in structures involving the quark vector $q$ and a new scalar source
\begin{equation}
	\mathcal{N}_1=
	\begin{pmatrix}
		\nu_1^{ub} & 0 & 0 \\
		0 & 0 & 0 \\
		0 & 0 & 0 
	\end{pmatrix}.
	\label{matrix2}
\end{equation}
We find the same transformation behaviour as for the 4q case. This leads to an
identical EFT Lagrangian construction procedure. Also here the 4qLR-terms
involving the constant $\nu_8^{ub}$ are not distinguishable from the $\nu_1^{ub}$-terms at the level
of chiral EFT. 
Finally, we mention that after rewriting the terms with Fierz identities,
we obtain both P- and T-violating and P- and T-conserving interactions. The latter lead to
modifications of P- and T-even observables that are swamped by Standard Model contributions, and
we neglect them below.

We are now in the position to write down the hadronic Lagrangians accounting for the various P- and
T-violating dimension six operators. For the quark EDM we obtain the two operators 
\begin{equation}                                
	\mathcal{L}_{qEDM}^{\rm eff.}=  c_1 \langle \bar{B}_{\bar{3}} \sigma^{\mu \nu} \gamma_5 F_{\mu \nu}  B_{\bar{3}} \rangle  +  c_2 \langle \bar{B}_{6} \sigma^{\mu \nu} \gamma_5 F_{\mu \nu}  B_{6} \rangle + \dots\,.
\end{equation}
A much longer list of operators appears for the qCEDM. Here, we give all operators that appear at the
same chiral order. As discussed below not all operators are relevant for the EDM calculations we perform.
We list them here for completeness. These read 
\begin{equation}
\begin{aligned} 
\mathcal{L}_{\mathrm{qCEDM}}^{\rm eff.}
=~& i \beta^+ \Big [ b_1 \langle \bar{B}_{\bar{3}} \chi_+ \gamma_5 B_{\bar{3}} \rangle + b_2 \langle \bar{B}_6 \chi_+ \gamma_5 B_6 \rangle + b_3 \langle \bar{B}_6 \chi_+ \gamma_5 B_{\bar{3}} + h.c.  \rangle + b_4 \langle \bar{B}_{\bar{3}} \gamma_5 B_{\bar{3}} \rangle \langle  \chi_+ \rangle  \\
                                             & + b_5 \langle \bar{B}_{6} \gamma_5 B_6 \rangle \langle  \chi_+ \rangle + b_6 \langle \bar{B}_{\bar{3}} \chi_- B_{\bar{3}} \rangle +b_7 \langle \bar{B}_6 \chi_- B_6 \rangle + b_8 \langle \bar{B}_6 \chi_- B_{\bar{3}} + h.c.  \rangle + b_9 \langle \bar{B}_{\bar{3}} B_{\bar{3}} \rangle \langle \chi_- \rangle  \\
                                             & + b_{10} \langle \bar{B}_{6} B_6 \rangle \langle \chi_- \rangle \Big ] + i \beta^+ \Big [ b_{11} \langle \bar{B}_{\bar{3}} u^{\mu} \gamma_5 u_{\mu} B_{\bar{3}} \rangle +b_{12} \langle \bar{B}_6 u^{\mu}  \gamma_5 u_{\mu} B_6 \rangle + b_{13} \langle \bar{B}_6 u^{\mu}  \gamma_5 u_{\mu} B_{\bar{3}} + h.c.  \rangle  \\  
                                             & + b_{14} \langle \bar{B}_{\bar{3}}  \gamma_5 B_{\bar{3}} \rangle \langle u^{\mu} u_{\mu} \rangle +b_{15} \langle \bar{B}_{6}  \gamma_5 B_6 \rangle \langle u^{\mu} u_{\mu} \rangle \Big] +  \beta^+ \Big [ b_{16} \langle \bar{B}_{\bar{3}} \sigma^{\mu \nu} \gamma_5 F_{\mu \nu}^+ B_{\bar{3}} \rangle +b_{17} \langle \bar{B}_6 \sigma^{\mu \nu} \gamma_5 F_{\mu \nu}^+ B_6 \rangle  \\
                                             & +b_{18} \langle \bar{B}_6 \sigma^{\mu \nu} \gamma_5 F_{\mu \nu}^+ B_{\bar{3}} + h.c.  \rangle
                                           + b_{19} \langle \bar{B}_{\bar{3}}\sigma^{\mu \nu} \gamma_5 B_{\bar{3}} \rangle \langle F_{\mu \nu}^+\rangle +b_{20} \langle \bar{B}_{6} \sigma^{\mu \nu} \gamma_5 B_6 \rangle \langle F_{\mu \nu}^+ \rangle \Big] \\
                                            & +  \beta^+ \Big[ b_{21} \langle \bar{B}_{\bar{3}} \sigma^{\mu \nu} \gamma_5 [u_{\mu}, u_{\nu}]B_{\bar{3}} \rangle + b_{22} \langle \bar{B}_{6} \sigma^{\mu \nu} \gamma_5 [u_{\mu}, u_{\nu}]B_6 \rangle + b_{23}\langle \bar{B}_{6} \sigma^{\mu \nu} \gamma_5 [u_{\mu}, u_{\nu}]B_{\bar{3}} + h.c. \rangle \Big] \\
                                            & +  \beta^+ \Big[ b_{24} \langle \bar{B}_{\bar{3}} u^{\mu} \rangle \langle u^{\nu} \sigma_{\mu \nu} \gamma_5 B_{\bar{3}} \rangle + b_{25} \langle \bar{B}_{6} u^{\mu} \rangle \langle u^{\nu} \sigma_{\mu \nu} \gamma_5 B_6 \rangle + b_{26} \langle \bar{B}_{6} u^{\mu} \rangle \langle u^{\nu} \sigma_{\mu \nu} \gamma_5 B_{\bar{3}} \rangle  + h.c.  \Big] \\
                                           & + i \beta^+ \Big[b_{27} \langle \bar{B}_{\bar{3}} u^{\mu} u^{\nu} \gamma_\mu \gamma_{5} D_\nu B_{\bar{3}} \rangle - b_{28} \langle \bar{B}_{\bar{3}} \overleftarrow{D}_{\nu} u^{\mu} u^{\nu} \gamma_\mu \gamma_{5} B_{\bar{3}} \rangle + b_{29} \langle \bar{B}_{6} u^{\mu} u^{\nu} \gamma_\mu \gamma_{5} D_\nu B_6 \rangle \\
                                           & - b_{30} \langle \bar{B}_{6} \overleftarrow{D}_{\nu} u^{\mu} u^{\nu} \gamma_\mu \gamma_{5} B_6 \rangle +b_{31} \langle \bar{B}_{6} u^{\mu} u^{\nu} \gamma_\mu \gamma_{5} D_\nu B_{\bar{3}} + h.c.  \rangle - b_{32} \langle \bar{B}_{6} \overleftarrow{D}_{\nu} u^{\mu} u^{\nu} \gamma_\mu \gamma_{5} B_{\bar{3}} + h.c.  \rangle  \Big]\\
                                          & + i \beta^+ \Big[\Big(b_{33} \langle \bar{B}_{\bar{3}} \gamma^\mu \gamma_{5} D^\nu B_{\bar{3}} \rangle - b_{34} \langle \bar{B}_{\bar{3}} \overleftarrow{D}^{\nu}\gamma^\mu \gamma_{5} B_{\bar{3}} \rangle \Big)  \langle u_{\mu} u_{\nu} \rangle + \Big( b_{35} \langle \bar{B}_{6}  \gamma^\mu \gamma_{5} D^\nu B_6 \rangle \\
                                           & - b_{36}\langle \bar{B}_{6} \overleftarrow{D}^{\nu} \gamma^\mu \gamma_{5} B_6 \rangle \Big) \langle u_{\mu} u_{\nu} \rangle + \Big(b_{37} \langle \bar{B}_{6} \gamma^\mu \gamma_{5} D^\nu B_{\bar{3}} + h.c.  \rangle -b_{38} \langle \bar{B}_{6} \overleftarrow{D}^{\nu} \gamma^\mu \gamma_{5} B_{\bar{3}} +h.c.  \rangle \Big)  \langle u_{\mu} u_{\nu} \rangle   \Big] \\
                                           & +\dots\,.
     \label{pt_vio_Lagr}                                        
	\end{aligned}                                          
\end{equation} 
	
For the four-quark operators we obtain 
\begin{equation}
	\begin{aligned}                                                                                      
	\mathcal{L}_{\mathrm{4q}}^{\rm eff.}=~& i\mu_1 \langle \bar{B}_{\bar{3}} \tilde{\chi}_+ \gamma_5 B_{\bar{3}} \rangle + i\mu_2 \langle \bar{B}_6 \tilde{\chi}_+ \gamma_5 B_6 \rangle +  i\mu_3 \langle \bar{B}_6 \tilde{\chi}_+ \gamma_5 B_{\bar{3}} + h.c.  \rangle + i\mu_4 \langle \bar{B}_{\bar{3}} \gamma_5 B_{\bar{3}} \rangle \langle  \tilde{\chi}_+ \rangle  \\
                                             &  + i\mu_5 \langle \bar{B}_{6} \gamma_5 B_6 \rangle \langle  \tilde{\chi}_+ \rangle + i\mu_6 \langle \bar{B}_{\bar{3}} \tilde{\chi}_- B_{\bar{3}} \rangle + i\mu_7 \langle \bar{B}_{6} \tilde{\chi}_- B_6 \rangle + i\mu_8 \langle \bar{B}_6 \tilde{\chi}_- B_{\bar{3}} + h.c.  \rangle \\
                                             & + i\mu_9 \langle \bar{B}_{\bar{3}}  B_{\bar{3}} \rangle  \langle \tilde{\chi}_- \rangle
                                          +i\mu_{10} \langle \bar{B}_{6}  B_{6} \rangle  \langle \tilde{\chi}_- \rangle + \mu_{11} \langle \bar{B}_{\bar{3}} \tilde{\chi}_+ \sigma^{\mu \nu} \gamma_5 F_{\mu \nu}^+ B_{\bar{3}} \rangle +\mu_{12} \langle \bar{B}_{6} \tilde{\chi}_+ \sigma^{\mu \nu} \gamma_5 F_{\mu \nu}^+ B_{6} \rangle \\
                                          & + \mu_{13} \langle \bar{B}_{6} \tilde{\chi}_+ \sigma^{\mu \nu} \gamma_5 F_{\mu \nu}^+ B_{\bar{3}} + h.c. \rangle + \mu_{14}  \langle \bar{B}_{\bar{3}} \tilde{\chi}_+ \sigma^{\mu \nu} \gamma_5 B_{\bar{3}} \rangle \langle F_{\mu \nu}^+ \rangle + \mu_{15} \langle \bar{B}_{6} \tilde{\chi}_+ \sigma^{\mu \nu} \gamma_5 B_{6} \rangle \langle F_{\mu \nu}^+ \rangle \\
                                          & + \mu_{16} \langle \bar{B}_{6} \tilde{\chi}_+ \sigma^{\mu \nu} \gamma_5 B_{\bar{3}} + h.c. \rangle \langle F_{\mu \nu}^+ \rangle + \mu_{17} \langle \bar{B}_{\bar{3}} \sigma^{\mu \nu} \gamma_5 F_{\mu \nu}^+ B_{\bar{3}} \rangle \langle \tilde{\chi}_+ \rangle + \mu_{18} \langle \bar{B}_{6} \sigma^{\mu \nu} \gamma_5 F_{\mu \nu}^+ B_{6} \rangle \langle \tilde{\chi}_+ \rangle \\ 
                                          & + \mu_{19} \langle \bar{B}_{6} \sigma^{\mu \nu} \gamma_5 F_{\mu \nu}^+ B_{\bar{3}} +h.c. \rangle \langle \tilde{\chi}_+ \rangle + \mu_{20} \langle \bar{B}_{\bar{3}} \sigma^{\mu \nu} \gamma_5  B_{\bar{3}} \rangle \langle \tilde{\chi}_+ F_{\mu \nu}^+ \rangle + \mu_{21} \langle \bar{B}_{6} \sigma^{\mu \nu} \gamma_5  B_{6} \rangle \langle \tilde{\chi}_+ F_{\mu \nu}^+ \rangle \\
                                          & + \mu_{22} \langle \bar{B}_{\bar{3}} \tilde{\chi}_- \sigma^{\mu \nu} F_{\mu \nu}^+ B_{\bar{3}} \rangle +\mu_{23} \langle \bar{B}_{6} \tilde{\chi}_- \sigma^{\mu \nu} F_{\mu \nu}^+ B_{6} \rangle 
                               				 + \mu_{24} \langle \bar{B}_{6} \tilde{\chi}_- \sigma^{\mu \nu}  F_{\mu \nu}^+ B_{\bar{3}} + h.c. \rangle \\
                               			& + \mu_{25}  \langle \bar{B}_{\bar{3}} \tilde{\chi}_- \sigma^{\mu \nu} B_{\bar{3}} \rangle \langle F_{\mu \nu}^+ \rangle + \mu_{26} \langle \bar{B}_{6} \tilde{\chi}_- \sigma^{\mu \nu} B_{6} \rangle \langle F_{\mu \nu}^+ \rangle 
                                          + \mu_{27} \langle \bar{B}_{6} \tilde{\chi}_- \sigma^{\mu \nu}  B_{\bar{3}} + h.c. \rangle \langle F_{\mu \nu}^+ \rangle \\
                                          & + \mu_{28} \langle \bar{B}_{\bar{3}} \sigma^{\mu \nu} F_{\mu \nu}^+ B_{\bar{3}} \rangle \langle \tilde{\chi}_- \rangle + \mu_{29} \langle \bar{B}_{6} \sigma^{\mu \nu} F_{\mu \nu}^+ B_{6} \rangle \langle \tilde{\chi}_- \rangle 
                                          + \mu_{30} \langle \bar{B}_{6} \sigma^{\mu \nu} F_{\mu \nu}^+ B_{\bar{3}} +h.c. \rangle \langle \tilde{\chi}_- \rangle \\
                                          & + \mu_{31} \langle \bar{B}_{\bar{3}} \sigma^{\mu \nu}  B_{\bar{3}} \rangle \langle \tilde{\chi}_- F_{\mu \nu}^+ \rangle + \mu_{32} \langle \bar{B}_{6} \sigma^{\mu \nu}  B_{6} \rangle \langle \tilde{\chi}_- F_{\mu \nu}^+ \rangle +\dots
                                          \end{aligned}         
                                          \end{equation}
                                          and
     \begin{equation}
	\begin{aligned}                                        
	\mathcal{L}_{\mathrm{4qLR}}^{\rm eff.}=~& i \text{Re} (V_{ub}) \Big[\nu_1 \langle \bar{B}_{\bar{3}} \hat{\tilde{\chi}}_-  B_{\bar{3}} \rangle + \nu_2 \langle \bar{B}_6 \hat{\tilde{\chi}}_- B_6 \rangle +  \nu_3 \langle \bar{B}_6 \hat{\tilde{\chi}}_- B_{\bar{3}} + h.c.  \rangle + \nu_4 \langle \bar{B}_{\bar{3}} B_{\bar{3}} \rangle \langle  \hat{\tilde{\chi}}_- \rangle \\
                                             &  + \nu_5 \langle \bar{B}_{6} B_6 \rangle \langle \hat{ \tilde{\chi}}_- \rangle + \nu_6 \langle \bar{B}_{\bar{3}} \hat{\tilde{\chi}}_+  \gamma_5 B_{\bar{3}} \rangle + \nu_7 \langle \bar{B}_{6} \hat{\tilde{\chi}}_+  \gamma_5  B_6 \rangle + \nu_8 \langle \bar{B}_6 \hat{\tilde{\chi}}_+ \gamma_5 B_{\bar{3}} + h.c.  \rangle  \\
                                            & + \nu_9 \langle \bar{B}_{\bar{3}}  \gamma_5  B_{\bar{3}} \rangle  \langle \hat{\tilde{\chi}}_+ \rangle
                                          + \nu_{10} \langle \bar{B}_{6}  \gamma_5  B_{6} \rangle  \langle \hat{\tilde{\chi}}_+ \rangle \Big] + \text{Re} (V_{ub}) \Big[ \nu_{11} \langle \bar{B}_{\bar{3}} \hat{\tilde{\chi}}_+ \sigma^{\mu \nu} \gamma_5 F_{\mu \nu}^+ B_{\bar{3}} \rangle \\ 
                                          & +\nu_{12} \langle \bar{B}_{6} \hat{\tilde{\chi}}_+ \sigma^{\mu \nu} \gamma_5 F_{\mu \nu}^+ B_{6} \rangle 
                                          + \nu_{13} \langle \bar{B}_{6} \hat{\tilde{\chi}}_+ \sigma^{\mu \nu} \gamma_5 F_{\mu \nu}^+ B_{\bar{3}} + h.c. \rangle + \nu_{14}  \langle \bar{B}_{\bar{3}} \hat{\tilde{\chi}}_+ \sigma^{\mu \nu} \gamma_5 B_{\bar{3}} \rangle \langle F_{\mu \nu}^+ \rangle \\
                                          & + \nu_{15} \langle \bar{B}_{6} \hat{\tilde{\chi}}_+ \sigma^{\mu \nu} \gamma_5 B_{6} \rangle \langle F_{\mu \nu}^+ \rangle 
                                          + \nu_{16} \langle \bar{B}_{6} \hat{\tilde{\chi}}_+ \sigma^{\mu \nu} \gamma_5 B_{\bar{3}} + h.c. \rangle \langle F_{\mu \nu}^+ \rangle + \nu_{17} \langle \bar{B}_{\bar{3}} \sigma^{\mu \nu} \gamma_5 F_{\mu \nu}^+ B_{\bar{3}} \rangle \langle \hat{\tilde{\chi}}_+ \rangle \\ 
                                          & + \nu_{18} \langle \bar{B}_{6} \sigma^{\mu \nu} \gamma_5 F_{\mu \nu}^+ B_{6} \rangle \langle \hat{\tilde{\chi}}_+ \rangle  
                                          + \nu_{19} \langle \bar{B}_{6} \sigma^{\mu \nu} \gamma_5 F_{\mu \nu}^+ B_{\bar{3}} +h.c. \rangle \langle \hat{\tilde{\chi}}_+ \rangle +\nu_{20} \langle \bar{B}_{\bar{3}} \sigma^{\mu \nu} \gamma_5  B_{\bar{3}} \rangle \langle \hat{\tilde{\chi}}_+ F_{\mu \nu}^+ \rangle \\
                                          & +\nu_{21} \langle \bar{B}_{6} \sigma^{\mu \nu} \gamma_5  B_{6} \rangle \langle \hat{\tilde{\chi}}_+ F_{\mu \nu}^+ \rangle + \nu_{22} \langle \bar{B}_{\bar{3}} \hat{\tilde{\chi}}_- \sigma^{\mu \nu} F_{\mu \nu}^+ B_{\bar{3}} \rangle  
                                          + \nu_{23} \langle \bar{B}_{6} \hat{\tilde{\chi}}_- \sigma^{\mu \nu} F_{\mu \nu}^+ B_{6} \rangle \\
                                          & + \nu_{24} \langle \bar{B}_{6} \hat{\tilde{\chi}}_- \sigma^{\mu \nu}  F_{\mu \nu}^+ B_{\bar{3}} + h.c. \rangle + \nu_{25}  \langle \bar{B}_{\bar{3}} \hat{\tilde{\chi}}_- \sigma^{\mu \nu} B_{\bar{3}} \rangle \langle F_{\mu \nu}^+ \rangle  
                                           + \nu_{26} \langle \bar{B}_{6} \hat{\tilde{\chi}}_- \sigma^{\mu \nu} B_{6} \rangle \langle F_{\mu \nu}^+ \rangle \\
                                          & + \nu_{27} \langle \bar{B}_{6} \hat{\tilde{\chi}}_- \sigma^{\mu \nu}  B_{\bar{3}} + h.c. \rangle \langle F_{\mu \nu}^+ \rangle + \nu_{28} \langle \bar{B}_{\bar{3}} \sigma^{\mu \nu} F_{\mu \nu}^+ B_{\bar{3}} \rangle \langle \hat{\tilde{\chi}}_- \rangle  
                                          + \nu_{29} \langle \bar{B}_{6} \sigma^{\mu \nu} F_{\mu \nu}^+ B_{6} \rangle \langle \hat{\tilde{\chi}}_- \rangle \\
                                          & + \nu_{30} \langle \bar{B}_{6} \sigma^{\mu \nu} F_{\mu \nu}^+ B_{\bar{3}} +h.c. \rangle \langle \hat{\tilde{\chi}}_- \rangle + \nu_{31} \langle \bar{B}_{\bar{3}} \sigma^{\mu \nu}  B_{\bar{3}} \rangle \langle \hat{\tilde{\chi}}_- F_{\mu \nu}^+ \rangle + \nu_{32} \langle \bar{B}_{6} \sigma^{\mu \nu}  B_{6} \rangle \langle \hat{\tilde{\chi}}_- F_{\mu \nu}^+ \rangle \Big] +\dots     \,.\label{pt_vio_Lagr}                                        
	\end{aligned}                                          
\end{equation} 
The ellipses indicate further terms of higher chiral order, which we will not display. 
We have defined 
\begin{equation}
	\begin{aligned}
\chi_{\pm}=~& u^{\dagger} \chi u^{\dagger} \pm u \chi ^{\dagger} u ,  \quad \quad \chi= 2 B_0\, \mathrm{diag}(m_u,\,m_d,\,m_s) , \\ 
	\tilde{\chi}_{\pm}= ~& u^{\dagger} \tilde{\chi} u^{\dagger} \pm u \tilde{\chi}^{\dagger} u ,  \quad \quad \tilde{\chi} \equiv   \text{diag} ~ (\mu^{ub},  \mu^{db},  \mu^{sb})\,,\\ 
	\hat{\tilde{\chi}}_{\pm} = ~& u^{\dagger} \hat{\tilde{\chi}} u^{\dagger} \pm u \hat{\tilde{\chi}}^{\dagger} u ,  \quad \quad \hat{\tilde{\chi}} \equiv   \text{diag} ~ ( \nu^{ub} ,0,0)\,, 
	\label{sources}
	\end{aligned}
\end{equation}
with the light quark masses $m_q$ and the LEC $B_0$ related to the quark condensate. 
Note that the constants $\mu^{ub}$, $\mu^{db}$, $\mu^{sb}$ and $\nu^{ub}$ capture both the color-singlet and -octet terms whose chiral Lagrangians are identical. 
	
It is convenient to use heavy-baryon chiral perturbation theory (HBChPT) while working with objects
that contain a single heavy quark \cite{Jenkins:1990jv,Bernard:1992qa}. In the heavy-baryon formulation,
several terms in the relativistic form cancel or appear at higher orders, and loop calculations
are simplified. Furthermore, the chiral power counting is manifest. 
The heavy-baryon Lagrangians are given by
\begin{equation}
	\begin{aligned}                     
	\mathcal{L}_{\text{free}}^{(1)}= ~&\frac{1}{2}\langle \bar{B}_{\bar{3},v}( i v\cdot D) B_{\bar{3},v} \rangle +\langle \bar{B}_{6,v}(i v \cdot D-\Delta) B_{6,v} \rangle,  \\                                                                                                                
	\mathcal{L}_{\text{int}}=~ ~ &g_1 \langle \bar{B}_{6,v} u_{\mu} S^{\mu} {B}_{6,v} \rangle+ g_2\langle \bar{B}_{6,v}u_{\mu} S^{\mu} {B}_{\bar{3},v}+h.c. \rangle , \\                                                                                                                                                    
	\mathcal{L}_{B \gamma}^{(2)}=~&2 \varepsilon^{\mu  \nu \rho \sigma} \Big[w_1 \langle \bar{B}_{\bar{3},v} v_{\rho} S_{\sigma} F_{\mu \nu}^+ B_{\bar{3},v} \rangle + w_2 \langle \bar{B}_{6,v} v_{\rho} S_{\sigma} F_{\mu \nu}^+ B_{6,v} \rangle + w_3 \langle \bar{B}_{6,v} v_{\rho} S_{\sigma} F_{\mu \nu}^+ B_{\bar{3},v} + h.c.  \rangle \\
                                               & + w_4 \langle \bar{B}_{\bar{3},v} v_{\rho} S_{\sigma}B_{\bar{3},v} \rangle \langle F_{\mu \nu}^+ \rangle + w_5 \langle \bar{B}_{6,v} v_{\rho} S_{\sigma} B_{6,v} \rangle \langle F_{\mu \nu}^+ \rangle \Big],
    \label{heavy Lagr}                                           
	\end{aligned}
\end{equation}
with the four velocity $v^{\mu}$, the Pauli-Lubanski spin operator $S^{\mu}=-\gamma_5(\gamma^{\mu}
\slashed{v}-v^{\mu})/2$ and the mass difference between sextet and anti-triplet baryons
$\Delta=m_6-m_{\bar{3}}$.  The effective Lagrangians for the P- and T-odd interactions
in the heavy baryon formulation are
\begin{equation}
\begin{aligned}
	\mathcal{L}_{\mathrm{qEDM}}^{\rm eff.}=~& 4 i \Big[ c_1 \langle \bar{B}_{\bar{3},v} v^{\mu} S^{\nu} F_{\mu \nu} B_{\bar{3},v} \rangle  + c_2 \langle \bar{B}_{6,v} v^{\mu} S^{\nu} F_{\mu \nu} B_{6,v} \rangle \Big], \\
		\mathcal{L}_{\mathrm{qCEDM}}^{\rm eff.}
	=~& 4 i \beta^+ \Big [ b_{16} \langle \bar{B}_{\bar{3},v} v^{\mu} S^{\nu} F_{\mu \nu}^+ B_{\bar{3},v} \rangle +b_{17} \langle \bar{B}_{6,v} v^{\mu} S^{\nu}  F_{\mu \nu}^+ B_{6,v} \rangle + b_{18} \langle \bar{B}_{6,v} v^{\mu} S^{\nu} F_{\mu \nu}^+ B_{\bar{3},v} + h.c.  \rangle \\
                                                & + b_{19} \langle \bar{B}_{\bar{3},v}v^{\mu} S^{\nu}  B_{\bar{3},v} \rangle \langle F_{\mu \nu}^+\rangle + b_{20} \langle \bar{B}_{6,v} v^{\mu} S^{\nu}  B_{6,v} \rangle \langle F_{\mu \nu}^+ \rangle \Big] + \dots \, ,  \\
     \mathcal{L}_{\mathrm{4q}}^{\rm eff.}=~& i\mu_6 \langle \bar{B}_{\bar{3},v} \tilde{\chi}_- B_{\bar{3},v} \rangle + i\mu_7 \langle \bar{B}_{6,v} \tilde{\chi}_- B_{6,v} \rangle + i\mu_8 \langle \bar{B}_{6,v} \tilde{\chi}_- B_{\bar{3},v} + h.c.  \rangle + i\mu_9 \langle \bar{B}_{\bar{3},v}  B_{\bar{3},v} \rangle  \langle \tilde{\chi}_- \rangle \\
                                                & +i\mu_{10} \langle \bar{B}_{6,v}  B_{6,v} \rangle  \langle \tilde{\chi}_- \rangle + 4 i \Big[ \mu_{11} \langle \bar{B}_{\bar{3},v} \tilde{\chi}_+ v^{\mu} S^{\nu} F_{\mu \nu}^+ B_{\bar{3},v} \rangle +\mu_{12} \langle \bar{B}_{6,v} \tilde{\chi}_+ v^{\mu} S^{\nu} F_{\mu \nu}^+ B_{6,v} \rangle \\
                                                & + \mu_{13} \langle \bar{B}_{6,v} \tilde{\chi}_+ v^{\mu} S^{\nu} F_{\mu \nu}^+ B_{\bar{3},v} + h.c. \rangle + \mu_{14}  \langle \bar{B}_{\bar{3},v} \tilde{\chi}_+ v^{\mu} S^{\nu} B_{\bar{3},v} \rangle \langle F_{\mu \nu}^+ \rangle + \mu_{15} \langle \bar{B}_{6,v} \tilde{\chi}_+ v^{\mu} S^{\nu} B_{6,v} \rangle \langle F_{\mu \nu}^+ \rangle \\
                                                & + \mu_{16} \langle \bar{B}_{6,v} \tilde{\chi}_+ v^{\mu} S^{\nu} B_{\bar{3},v} + h.c. \rangle \langle F_{\mu \nu}^+ \rangle + \mu_{17} \langle \bar{B}_{\bar{3},v} v^{\mu} S^{\nu} F_{\mu \nu}^+ B_{\bar{3},v} \rangle \langle \tilde{\chi}_+ \rangle + \mu_{18} \langle \bar{B}_{6,v} v^{\mu} S^{\nu} F_{\mu \nu}^+ B_{6,v} \rangle \langle \tilde{\chi}_+ \rangle \\ 
                                                & + \mu_{19} \langle \bar{B}_{6,v} v^{\mu} S^{\nu} F_{\mu \nu}^+ B_{\bar{3},v} +h.c. \rangle \langle \tilde{\chi}_+ \rangle + \mu_{20} \langle \bar{B}_{\bar{3},v} v^{\mu} S^{\nu} B_{\bar{3},v} \rangle \langle \tilde{\chi}_+ F_{\mu \nu}^+ \rangle + \mu_{21} \langle \bar{B}_{6,v} v^{\mu} S^{\nu} B_{6,v} \rangle \langle \tilde{\chi}_+ F_{\mu \nu}^+ \rangle \Big] \\ 
                                                & + \dots \, , \\
\end{aligned}
\end{equation}

\begin{equation}
	\begin{aligned}                                                                
	\mathcal{L}_{\mathrm{4qLR}}^{\rm eff.}=~& i \text{Re} (V_{ub}) \Big[\nu_1  \langle \bar{B}_{\bar{3},v} \hat{\tilde{\chi}}_-  B_{\bar{3},v} \rangle + \nu_2 \langle \bar{B}_{6,v} \hat{\tilde{\chi}}_- B_{6,v} \rangle +  \nu_3 \langle \bar{B}_{6,v} \hat{\tilde{\chi}}_- B_{\bar{3},v} + h.c.  \rangle + \nu_4 \langle \bar{B}_{\bar{3},v} B_{\bar{3},v} \rangle \langle  \hat{\tilde{\chi}}_- \rangle \\
                                         	 & + \nu_5 \langle \bar{B}_{6,v} B_{6,v} \rangle \langle \hat{ \tilde{\chi}}_- \rangle \Big] + 4 i \text{Re} (V_{ub}) \Big[ \nu_{11} \langle \bar{B}_{\bar{3},v} \hat{\tilde{\chi}}_+ v^{\mu} S^{\nu} F_{\mu \nu}^+ B_{\bar{3},v} \rangle +\nu_{12} \langle \bar{B}_{6,v} \hat{\tilde{\chi}}_+ v^{\mu} S^{\nu} F_{\mu \nu}^+ B_{6,v} \rangle \\
                                         	 & + \nu_{13} \langle \bar{B}_{6,v} \hat{\tilde{\chi}}_+ v^{\mu} S^{\nu} F_{\mu \nu}^+ B_{\bar{3},v} + h.c. \rangle + \nu_{14}  \langle \bar{B}_{\bar{3},v} \hat{\tilde{\chi}}_+ v^{\mu} S^{\nu} B_{\bar{3},v} \rangle \langle F_{\mu \nu}^+ \rangle  + \nu_{15} \langle \bar{B}_{6,v} \hat{\tilde{\chi}}_+ v^{\mu} S^{\nu} B_{6,v} \rangle \langle F_{\mu \nu}^+ \rangle \\
                                         	 & + \nu_{16} \langle \bar{B}_{6,v} \hat{\tilde{\chi}}_+ v^{\mu} S^{\nu} B_{\bar{3},v} + h.c. \rangle \langle F_{\mu \nu}^+ \rangle + \nu_{17} \langle \bar{B}_{\bar{3},v} v^{\mu} S^{\nu} F_{\mu \nu}^+ B_{\bar{3},v} \rangle \langle \hat{\tilde{\chi}}_+ \rangle  + \nu_{18} \langle \bar{B}_{6,v} v^{\mu} S^{\nu} F_{\mu \nu}^+ B_{6,v} \rangle \langle \hat{\tilde{\chi}}_+ \rangle \\
                                         	 & + \nu_{19} \langle \bar{B}_{6,v} v^{\mu} S^{\nu} F_{\mu \nu}^+ B_{\bar{3},v} +h.c. \rangle \langle \hat{\tilde{\chi}}_+ \rangle + \nu_{20} \langle \bar{B}_{\bar{3},v} v^{\mu} S^{\nu} B_{\bar{3},v} \rangle \langle \hat{\tilde{\chi}}_+ F_{\mu \nu}^+ \rangle + \nu_{21} \langle \bar{B}_{6,v} v^{\mu} S^{\nu} B_{6,v} \rangle \langle \hat{\tilde{\chi}}_+ F_{\mu \nu}^+ \rangle \Big] \\
                                         	 & + \dots \, .                                  
	\end{aligned}                                          
\end{equation}
We only display the terms which are relevent for the EDM calculation.  Additionally, to the order we
are working only terms linear in the Goldstone bosons are needed.  Terms that begin with more than a single Goldstone boson are hidden in the ellipses. Since the chiral singlet $\beta^+$ in the qCEDM Lagrangian can only enter as an overall constant, it is convenient to absorb $\beta^+$ into the LECs $b_i$.

Concerning the power counting rules of the CP-odd vertices, the chiral order of the sources is
counted as $\mathcal{O}(\delta^0)$, where $\delta$ is a generic small mass or momentum,
since they do not contain any light scales and in addition will be common to all
contributions considered in this work. For the remaining pieces  we employ standard chiral
counting. 

\begin{figure}[h!]
	\centering
	\includegraphics[width=0.8\textwidth]{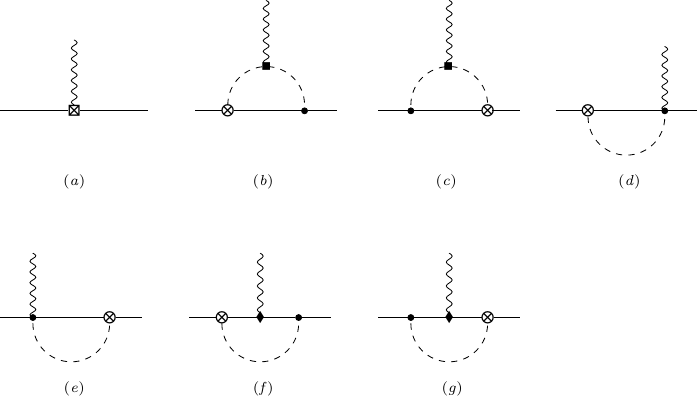}
	\caption{Diagrams contributing to the EDMs of the spin-$1/2$ neutral anti-triplet and sextet
          $b$-baryons. Solid lines correspond to contribution from either spin-$1/2$ anti-triplet
          or sextet multiplets of bottom baryons.
	  Filled circles and squares are first-order meson-baryon and second order mesonic vertices,
          respectively. While diamonds represent vertices generated by the first order
          meson-baryon Lagrangian, CP-violating vertices at $\mathcal{O}(\delta^0)$
          and $\mathcal{O}(\delta^2)$ are represented by $\otimes$ and $\boxtimes$, in order.}
	\label{fig:diag}
\end{figure}

Fig.~\ref{fig:diag} depicts the tree-level and one-loop Feynman
diagrams that generate a non-vanishing contribution to the P- and T-violating form factor
of the $B_b$ baryons up to the order $\mathcal{O}(\delta^2)$.
We evaluate the loop diagrams in the framework of dimensional
regularization at the renormalization scale $\lambda=1$~GeV. 
We apply the modified minimal subtraction scheme ($\widetilde{\text{MS}}$) in HBChPT 
\cite{tHooft:1973mfk, Weinberg:1973xwm,Gasser:1983yg, Scherer:2012zzd} by absorbing the
infinite parts in terms of 
\begin{eqnarray}
 L =~\frac{\lambda^{n-4}}{16 \pi^2}\,\Bigg[\frac{1}{n-4}+\frac{1}{2}\Big(\gamma_E-1-\text{ln}(4 \pi)\Big)\Bigg].
\end{eqnarray}
into the counterterms, with $n$ the number of space-time dimensions and $\gamma_E$ the Euler-Mascheroni constant.
The tree-level CP-odd diagrams at order $\mathcal{O}(\delta^2)$
displayed in diagram (a) receive contributions from all the CP-violating operators. 
The one-loop diagrams at leading $\mathcal{O}(\delta^2)$ are given by diagrams $(b)$-$(g)$ in
Fig.~\ref{fig:diag}.  

\section{The P- and T-violating form factor}\label{sec:loop}

The EDM of the neutral and charged $b$-baryons can
be extracted from the P- and T-violating form factor $D_{B_b}^{\gamma}(q^2)$.
It is defined through
\begin{eqnarray}
 \Braket{B_b (p_{f})|J_{\text{EDM}, \nu}|B_b (p_i)}
    = ~ D_{B_b}^{\gamma}(q^2)\,\bar{u}(p_f) \sigma_{\mu \nu} \gamma_5 q^{\mu} u(p_i)\,,
\end{eqnarray}	
in the covariant formulation with momentum transfer $q = p_f-p_i$, see e.g. Ref.~\cite{Borasoy:2000pq}.
The EDM is then given by 
\begin{eqnarray}
d_{B_b}^\gamma = D_{B_b}^{\gamma}(q^2=0).
\end{eqnarray}	
One can reformulate the form factor in the heavy baryon approach using the Breit frame.
In this frame, we have $v\cdot p_i=v\cdot p_f$ and we set the four-velocity to
$v_{\mu}=(1,\textbf{0})$. The form factor is then obtained as 
\begin{eqnarray}
 \Braket{B_{b} (p_{f})|J_{\text{EDM}, \nu}|B_{b} (p_i)} = -2 i D_{B_b}^{\gamma}(q^2) \bar{B}_{v} v_{\nu} (S \cdot q) B_{v} .
\end{eqnarray}	
We first consider the contributions from the tree-level diagrams in Fig.~\ref{fig:diag}-(a).
The expressions of the electric dipole moment of the anti-triplet and sextet $b$-baryons from the
dimension six operators are collected in Tables \ref{tab:tree1}-\ref{tab:tree3}.
In addition to the tree-level contributions, we find the one-loop diagrams in Fig.~\ref{fig:diag}.
In analogy to the neutron EDM, the EDMs of bottom baryons get contributions from the cloud of
Goldstone bosons dressing the baryons. 
For the qEDM and qCEDM operators the meson-loops appear at higher order and only the tree-level diagrams
are necessary. But for the 4q and 4qLR operators the loops appear at the same order and the LECs of
the tree-level contributions absorb the associated loop divergences. 

We calculated the diagrams in Fig.~\ref{fig:diag} explicitly in heavy-baryon ChPT. We find that only
diagrams $(b)$ and $(c)$ contribute at the order we work. 
The other diagrams are proportional to $S \cdot v =0$, or $v\cdot q=0$, or mutually cancel.
The contributions from the non-vanishing diagrams can be written as
\begin{equation}
	\begin{aligned} 
	D^{\gamma}_b(q^2)= ~&\frac{A_{b_i}}{2} \int_0^1 dx \, \frac{x}{\tilde{M}_i}\frac{\partial}{\partial \tilde{M}_i}J_1(\tilde{w}, \tilde{M}_i), \\
	D^{\gamma}_c(q^2)=  ~&\frac{A_{c_i}}{2} \int_0^1 dx \, \frac{x-1}{\tilde{M}_i}\frac{\partial}{\partial \tilde{M}_i}J_1(\tilde{w}, \tilde{M}_i),  \quad i = 1, 2, 3, 4.
     \label{res4q-edm}
	\end{aligned}
\end{equation} 
where $J_1$ is the loop function defined in App.~\ref{sec:AppendixC}, $\tilde{w}=-\Delta$ for a
sextet particle inside the loop, or $\tilde{w}=0$ for an anti-triplet particle. Furthermore,
$\tilde{M}_i(x)=\sqrt{x(x-1)q^2+M_i^2}$, with $M_i$ being $M_{K^{\pm}}$ or $M_{\pi^{\pm}}$. The coefficients
$A_{b_i}$ and $A_{c_i}$ have to be determined from the vertices of the appearing interacting Lagrangians.
A lot of these coefficients are similar to each other with some only differing by their sign.
Considering  isospin symmetry this leads to additional cancellations when summing up the loop contributions.
We refrain from showing the full list of coefficients $A_{b_i}$ and $A_{c_i}$ with their respective $M_i$ here.
The surviving terms together with their coefficients can be read off from the full form factor
results listed in App.~\ref{sec:AppendixA}.  

 \begin{table}[h!]
 	\centering
	\caption{Tree-level contributions from the qEDM and qCEDM operators of the $b$-baryons.
          Loop diagrams only appear at higher order.}
	\begin{tabular}{ccc}
		\toprule
{Baryons}                   &  {qEDM}          & {qCEDM}     \\
		\midrule  
{$\Lambda_{b}^0$}  
  & {$ 4 c_1$} 
  &  {$-4e  b_{19}$} \\
{$\Xi_{b}^0$}           
  & {$ 4 c_1$} 
  &  {$-4e  b_{19}$} \\
 {$\Xi_{b}^-$}          
  & {$ 4 c_1$} 
  & {$-4e  (b_{16} + b_{19})$} \\
		\midrule
 {$\Sigma_{b}^{+}$}
  & {$ 2 c_2$} 
  &  {$2e  (b_{17}-b_{20})$} \\
{$\Sigma_{b}^{0}$}       
  & {$ 2 c_2$} 
  &  {$-2e  b_{20}$} \\
{$\Sigma_{b}^{-}$}       
  & {$ 2 c_2$} 
  &  {$-2e  (b_{17}+b_{20})$} \\
{$\Xi_{b}^{'0}$}              
  & {$ 2 c_2$} 
  &  {$-2e  b_{20}$} \\
{$\Xi_{b}^{'-}$}              
  & {$ 2 c_2$} 
  &  {$-2e (b_{17} + b_{20})$} \\
{$\Omega_{b}^{-}$}      
  & {$ 2 c_2 $} 
  &  {$-2e  (b_{17} + b_{20})$} \\
		\bottomrule
	\end{tabular}
	\label{tab:tree1}
\end{table}

 \begin{table}[h!]
 	\centering
	\caption{Tree-level contribution from the 4q operators of the $b$-baryons. Loop diagrams
          appear at the same order. } 
\begin{tabular}{cc}
		\toprule
{Baryons}             & {4q}               \\
		\midrule  
{$\Lambda_{b}^0$}  
  & {$ 4e[\mu_{11}(\mu^{ub}-\mu^{db})-\mu_{14}(\mu^{ub}+\mu^{db}) +2 \mu_{20} (\mu^{ub}-\mu^{db}-\mu^{sb})]$} 
  \\
{$\Xi_{b}^0$}           
  & {$4e[\mu_{11}(\mu^{ub}-\mu^{sb})-\mu_{14}(\mu^{ub}+\mu^{sb})+2 \mu_{20} (\mu^{ub}-\mu^{db}-\mu^{sb})]$}  \\
  
{$\Xi_{b}^-$}          
  & {$-4e[(\mu_{11}+\mu_{14})(\mu^{db}+\mu^{sb})+2 \mu_{17}(\mu^{ub}+\mu^{db}+\mu^{sb})-2 \mu_{20} (\mu^{ub}-\mu^{db}-\mu^{sb})]$} \\
  
		\midrule
{$\Sigma_{b}^{+}$}
  & {$ 4e [(\mu_{12}-\mu_{15}) \mu^{ub}+\mu_{18}(\mu^{ub}+\mu^{db}+\mu^{sb}) + \mu_{21} (\mu^{ub}-\mu^{db}-\mu^{sb})]$} \\
  
{$\Sigma_{b}^{0}$}       
  & {$ 2e [\mu_{12} (\mu^{ub}-\mu^{db})-\mu_{15}(\mu^{ub}+\mu^{db})+2\mu_{21} (\mu^{ub}-\mu^{db}-\mu^{sb})]$} \\
  
{$\Sigma_{b}^{-}$}       
  & {$-4e[(\mu_{12}+\mu_{15})\mu^{db}+\mu_{18}(\mu^{ub}+\mu^{db}+\mu^{sb})-\mu_{21} (\mu^{ub}-\mu^{db}-\mu^{sb})]$} \\
  
{$\Xi_{b}^{'0}$}              
  & {$ 2e[\mu_{12}(\mu^{ub}-\mu^{sb})-\mu_{15}(\mu^{ub}+\mu^{sb})+2\mu_{21} (\mu^{ub}-\mu^{db}-\mu^{sb})]$} \\
  
{$\Xi_{b}^{'-}$}              
  & {$-2e[(\mu_{12}+\mu_{15})(\mu^{db}+\mu^{sb})+2 \mu_{18}(\mu^{ub}+\mu^{db}+\mu^{sb})-2\mu_{21} (\mu^{ub}-\mu^{db}-\mu^{sb})]$} \\
 
{$\Omega_{b}^{-}$}      
  & {$-4e[(\mu_{12}+\mu_{15})\mu^{sb}+\mu_{18}(\mu^{ub}+\mu^{db}+\mu^{sb})-\mu_{21} (\mu^{ub}-\mu^{db}-\mu^{sb})]$} \\
		\bottomrule
	\end{tabular}
	\label{tab:tree2}
\end{table}

 \begin{table}[h!]
	\centering
	\caption{Tree-level contribution from 4qLR operators of the $b$-baryons. Loop diagrams appear
          at the same order. } 
	\begin{tabular}{cc}
		\toprule
		{Baryons}                          & {4qLR}   \\
		\midrule  
		{$\Lambda_{b}^0$}  
		
		& {$ 4e \text{Re}(V_{ub})(\nu_{11}-\nu_{14} + 2 \nu_{20}) \nu^{ub} $} \\
		{$\Xi_{b}^0$}           
		
		& {$4e \text{Re}(V_{ub})(\nu_{11}-\nu_{14} + 2 \nu_{20})\nu^{ub}$}  \\
		{$\Xi_{b}^-$}          
		
		& {$-8e \text{Re}(V_{ub})(\nu_{17}-\nu_{20})\nu^{ub}$} \\
		\midrule
		{$\Sigma_{b}^{+}$}
		& {$ 4e \text{Re}(V_{ub})(\nu_{12}-\nu_{15}+\nu_{18}+\nu_{21})\nu^{ub}$} \\
		{$\Sigma_{b}^{0}$}       

		& {$ 2e \text{Re}(V_{ub})(\nu_{12}-\nu_{15}+2\nu_{21})\nu^{ub}$} \\
		{$\Sigma_{b}^{-}$}       

		& {$ -4e \text{Re}(V_{ub})(\nu_{18}-\nu_{21})\nu^{ub}$} \\
		{$\Xi_{b}^{'0}$}              

		& {$ 2 e \text{Re}(V_{ub})(\nu_{12}-\nu_{15}+2\nu_{21})\nu^{ub}$} \\
		{$\Xi_{b}^{'-}$}              

		& {$ - 4e \text{Re}(V_{ub})(\nu_{18}-\nu_{21})\nu^{ub}$} \\
		{$\Omega_{b}^{-}$}      

		& {$ -4e \text{Re}(V_{ub})(\nu_{18}-\nu_{21})\nu^{ub}$} \\
		\bottomrule
	\end{tabular}
	\label{tab:tree3}
\end{table}

The expressions for the complete form factors are given in App.~\ref{sec:AppendixA}. Here, we present
the results for the EDMs. For the 4q operator we obtain

\begin{equation}
	\begin{aligned} 
		d^{\gamma}_{\Lambda_b^0, 4q}= ~& 4e\Big [\mu_{11}(\mu^{ub}-\mu^{db})-\mu_{14}(\mu^{ub}+\mu^{db}) +2 \mu_{20} (\mu^{ub}-\mu^{db}-\mu^{sb}) \Big] + \frac{e g_2 \mu_8 ( \mu^{ub}+\mu^{sb})}{32 \pi^2 F_{\pi}^2} F_{M_K}^{(2)} ,\\                                  
		d^{\gamma}_{\Xi_b^0,\mathrm{4q}}= ~& 4e\Big [\mu_{11}(\mu^{ub}-\mu^{sb})-\mu_{14}(\mu^{ub}+\mu^{sb}) +2 \mu_{20} (\mu^{ub}-\mu^{db}-\mu^{sb}) \Big] + \frac{e g_2  \mu_8 ( \mu^{ub}+\mu^{db})}{32 \pi^2 F_{\pi}^2}  F_{M_\pi}^{(2)}, \\
		    d^{\gamma}_{\Xi_b^-,\mathrm{4q}}= ~&-4e \Big [(\mu_{11}+\mu_{14})(\mu^{db}+\mu^{sb})+2 \mu_{17}(\mu^{ub}+\mu^{db}+\mu^{sb}) -2 \mu_{20} (\mu^{ub}-\mu^{db}-\mu^{sb}) \Big] \\
		& - \frac{e g_2  \mu_8} {32 \pi^2 F_{\pi}^2} \Bigg ( (\mu^{ub}+\mu^{sb})F_{M_K}^{(2)} +
		 (\mu^{ub}+\mu^{db}) F_{M_\pi}^{(2)} \Bigg), 
		\label{all-edm-4q}
	\end{aligned}
\end{equation} 

\begin{equation}
	\begin{aligned}                  
	     d^{\gamma}_{\Sigma_b^+,\mathrm{4q}}= ~& 4e \Big [(\mu_{12}-\mu_{15}) \mu^{ub}+\mu_{18}(\mu^{ub}+\mu^{db}+\mu^{sb}) + \mu_{21} (\mu^{ub}-\mu^{db}-\mu^{sb}) \Big] \\
		& + \frac{e g_1  \mu_7}{32 \pi^2 F_{\pi}^2} \Bigg ( (\mu^{ub}+\mu^{sb}) F_{M_K}^{(2)}
		 + ( \mu^{ub}+\mu^{db}) F_{M_\pi}^{(2)} \Bigg) \\
		& + \frac{e g_2 \mu_8}{16 \pi^2 F_{\pi}^2} \Bigg( ( \mu^{ub}+\mu^{sb}) F_{M_K}^{(1)} 
		 +( \mu^{ub}+\mu^{db}) F_{M_\pi}^{(1)} \Bigg) , \\
		 d^{\gamma}_{\Sigma_b^0,\mathrm{4q}}=  ~& 2e \Big [\mu_{12} (\mu^{ub}-\mu^{db})-\mu_{15}(\mu^{ub}+\mu^{db}) +2 \mu_{21} (\mu^{ub}-\mu^{db}-\mu^{sb}) \Big] \\
		   &  + \frac{e g_1 \mu_7 ( \mu^{ub}+\mu^{sb})}{64 \pi^2 F_{\pi}^2}  F_{M_K}^{(2)}   
	         + \frac{e g_2 \mu_8 ( \mu^{ub}+\mu^{sb})}{32 \pi^2 F_{\pi}^2}  F_{M_K}^{(1)} ,  \\          		 		   	     
	      d^{\gamma}_{\Sigma_b^-,\mathrm{4q}}=  ~& -4e \Big[(\mu_{12}+\mu_{15})\mu^{db}+\mu_{18}(\mu^{ub}+\mu^{db}+\mu^{sb}) - \mu_{21} (\mu^{ub}-\mu^{db}-\mu^{sb}) \Big] \\
		& - \frac{e g_1 \mu_7 ( \mu^{ub}+\mu^{db})}{32 \pi^2 F_{\pi}^2} 
		 F_{M_\pi}^{(2)}  
		 - \frac{e g_2 \mu_8 ( \mu^{ub}+\mu^{db}) }{16 \pi^2 F_{\pi}^2}  F_{M_\pi}^{(1)} ,    \\                      
		d^{\gamma}_{\Xi_b^{'0},\mathrm{4q}}=  ~&  2e \Big [\mu_{12}(\mu^{ub}-\mu^{sb})-\mu_{15}(\mu^{ub}+\mu^{sb}) +2 \mu_{21} (\mu^{ub}-\mu^{db}-\mu^{sb})\Big] \\
		    & + \frac{e g_1 \mu_7 ( \mu^{ub}+\mu^{db})}{64 \pi^2 F_{\pi}^2} 
		    F_{M_\pi}^{(2)} 
		     + \frac{e g_2 \mu_8 ( \mu^{ub}+\mu^{db})}{32 \pi^2 F_{\pi}^2} \ F_{M_\pi}^{(1)} ,  \\
		     d^{\gamma}_{\Xi_b^{'-},\mathrm{4q}}=  ~&  -2e \Big[(\mu_{12}+\mu_{15})(\mu^{db}+\mu^{sb})+2 \mu_{18}(\mu^{ub}+\mu^{db}+\mu^{sb}) -2 \mu_{21} (\mu^{ub}-\mu^{db}-\mu^{sb}) \Big]\\
		& - \frac{e g_1 \mu_7} {64 \pi^2 F_{\pi}^2} \Bigg( (\mu^{ub}+\mu^{sb})  F_{M_K}^{(2)} 
		 + ( \mu^{ub}+\mu^{db}) F_{M_\pi}^{(2)} \Bigg) \\
		& - \frac{e g_2 \mu_8} {32 \pi^2 F_{\pi}^2} \Bigg ( ( \mu^{ub}+\mu^{sb}) F_{M_K}^{(1)} 
		 +( \mu^{ub}+\mu^{db}) F_{M_\pi}^{(1)} \Bigg) ,\\	
	d^{\gamma}_{\Omega_b^{-},\mathrm{4q}}=  ~& -4e\Big[(\mu_{12}+\mu_{15})\mu^{sb}+\mu_{18}(\mu^{ub}+\mu^{db}+\mu^{sb}) - \mu_{21} (\mu^{ub}-\mu^{db}-\mu^{sb}) \Big] \\
		& - \frac{e g_1 \mu_7 ( \mu^{ub}+\mu^{sb}) }{32 \pi^2 F_{\pi}^2}  F_{M_K}^{(2)} 
		 - \frac{e g_2 \mu_8 ( \mu^{ub}+\mu^{sb})}{16 \pi^2 F_{\pi}^2}  F_{M_K}^{(1)} \,.
		\label{all-edm-4q}
	\end{aligned}
\end{equation} 
For the 4qLR operator we obtain
\begin{equation}
	\begin{aligned} 
		d^{\gamma}_{\Lambda_b^0, 4qLR}= ~& 4e \text{Re}(V_{ub})(\nu_{11}-\nu_{14} + 2 \nu_{20}) \nu^{ub} 
			 + \frac{e \text{Re}(V_{ub}) g_2  \nu_3 \nu^{ub}}{32 \pi^2 F_{\pi}^2} F_{M_K}^{(2)} ,
			 \\
		d^{\gamma}_{\Xi_b^0,\mathrm{4qLR}}= ~& 4e \text{Re}(V_{ub})(\nu_{11}-\nu_{14} + 2 \nu_{20}) \nu^{ub}
			 + \frac{e \text{Re}(V_{ub})g_2 \nu_3 \nu^{ub} }{32 \pi^2 F_{\pi}^2} 
		    F_{M_\pi}^{(2)} ,
		     \\
		    d^{\gamma}_{\Xi_b^-,\mathrm{4qLR}}= ~&-8e \text{Re}(V_{ub})(\nu_{17}-\nu_{20}) \nu^{ub}
		 - \frac{e \text{Re}(V_{ub})g_2 \nu_3 \nu^{ub}} {32 \pi^2 F_{\pi}^2} \Bigg(F_{M_K}^{(2)} 
		 + F_{M_\pi}^{(2)} \Bigg),  
		    \\   
       d^{\gamma}_{\Sigma_b^+,\mathrm{4qLR}}= ~& 4e \text{Re}(V_{ub})(\nu_{12}-\nu_{15}+\nu_{18}+\nu_{21}) \nu^{ub} \\
		& + \frac{e \text{Re}(V_{ub}) g_1 \nu_2 \nu^{ub}}{32 \pi^2 F_{\pi}^2} \Big( F_{M_K}^{(2)} +
		F_{M_\pi}^{(2)} \Big) 
		+ \frac{e \text{Re}(V_{ub}) g_2 \nu_3 \nu^{ub}}{16 \pi^2 F_{\pi}^2} \Big(F_{M_K}^{(1)} 
		+ F_{M_\pi}^{(1)} \Big) ,                    		 		   	     
		\label{all-edm-4qLR}
	\end{aligned}
\end{equation} 

\begin{equation}
	\begin{aligned} 
	d^{\gamma}_{\Sigma_b^0,\mathrm{4qLR}}=  ~& 2e \text{Re}(V_{ub})(\nu_{12}-\nu_{15}+ 2\nu_{21}) \nu^{ub} 
			 + \frac{e \text{Re}(V_{ub})g_1 \nu_2 \nu^{ub}}{64 \pi^2 F_{\pi}^2}F_{M_K}^{(2)} 
	        + \frac{e \text{Re}(V_{ub})g_2 \nu_3 \nu^{ub}}{32 \pi^2 F_{\pi}^2} F_{M_K}^{(1)} ,
	     \\
	      d^{\gamma}_{\Sigma_b^-,\mathrm{4qLR}}=  ~&  -4e \text{Re}(V_{ub})(\nu_{18}- \nu_{21}) \nu^{ub} 
		 - \frac{e \text{Re}(V_{ub}) g_1 \nu_2 \nu^{ub}}{32 \pi^2 F_{\pi}^2} 
		 F_{M_\pi}^{(2)} 
		  - \frac{e  \text{Re}(V_{ub})g_2 \nu_3 \nu^{ub}}{16 \pi^2 F_{\pi}^2} F_{M_\pi}^{(1)} ,                          
          \\                       
		d^{\gamma}_{\Xi_b^{'0},\mathrm{4qLR}}=  ~&  2 e \text{Re}(V_{ub})(\nu_{12}-\nu_{15}+ 2\nu_{21}) \nu^{ub} 
			+ \frac{e \text{Re}(V_{ub}) g_1 \nu_2 \nu^{ub}}{64 \pi^2 F_{\pi}^2} 
		   F_{M_\pi}^{(2)} 
		     + \frac{e \text{Re}(V_{ub}) g_2 \nu_3 \nu^{ub}}{32 \pi^2 F_{\pi}^2} F_{M_\pi}^{(1)} ,           
	   \\                        
		d^{\gamma}_{\Xi_b^{'-},\mathrm{4qLR}}=  ~& - 4e \text{Re}(V_{ub})(\nu_{18} - \nu_{21}) \nu^{ub}
		 - \frac{e \text{Re}(V_{ub}) g_1 \nu_2 \nu^{ub}}{64 \pi^2 F_{\pi}^2} \Big(F_{M_K}^{(2)}  
	      +  F_{M_\pi}^{(2)} \Big) \\
		 & - \frac{e \text{Re}(V_{ub}) g_2 \nu_3 \nu^{ub}}{32 \pi^2 F_{\pi}^2} \Big(F_{M_K}^{(1)}  
		 + F_{M_\pi}^{(1)} \Big) ,
		\\	
	d^{\gamma}_{\Omega_b^{-},\mathrm{4qLR}}=  ~& -4e \text{Re}(V_{ub})(\nu_{18} - \nu_{21}) \nu^{ub} 
		 - \frac{e \text{Re}(V_{ub}) g_1 \nu_2 \nu^{ub}}{32 \pi^2 F_{\pi}^2} F_{M_K}^{(2)} 
		 - \frac{e \text{Re}(V_{ub}) g_2 \nu_3 \nu^{ub}}{16 \pi^2 F_{\pi}^2} F_{M_K}^{(1)} \, ,	
	  \label{all-edm-4qLR}
	\end{aligned}
\end{equation} 
where the loop functions are defined as 
\begin{equation}
      \begin{aligned}
F_{M_\pi}^{(1)} = ~& 1 + 32 \pi^2 L + 2 \text{ln}\Bigg[\frac{M_\pi}{\lambda} \Bigg] ,  \\
F_{M_\pi}^{(2)} = ~& 1 + 32 \pi^2 L + 2 \text{ln}\Bigg[\frac{M_{\pi}}{\lambda}\Bigg] + \frac{2 \Delta}{\sqrt{\Delta^2 - M_{\pi}^2 }} \text{ln}\Bigg[ \frac{\Delta}{M_{\pi}}  + \sqrt{\frac{\Delta^2}{M_{\pi}^2} -1 } \Bigg] , \\
F_{M_K}^{(1)} = ~& 1 + 32 \pi^2 L + 2 \text{ln}\Bigg[\frac{M_K}{\lambda} \Bigg] , \\
F_{M_K}^{(2)} = ~& 1 + 32 \pi^2 L + 2 \text{ln}\Bigg[\frac{M_K}{\lambda}\Bigg] + \frac{2 \Delta {\arccos}\big[\frac{\Delta}{M_K}\big]}{\sqrt{M_K^2-\Delta^2}}\,.
     \end{aligned}
\end{equation}

\subsection{Patterns of EDMs}
Before discussing the absolute sizes of the EDMs in the next section, we investigate the relative
sizes of the various EDMs. The relative sizes are essentially determined by the chiral symmetry
properties and field content of the underlying sources of CP violation. For instance, for the bottom qEDM at
the order at which we work, the EDMs of all baryons in the triplet or the sextet are determined by a single
LEC, $c_1$ and $c_2$, in order. This pattern is different for the qCEDM where in the triplet $d_{\Xi_b^-}$ is
expected to be different from $d^\gamma_{\Lambda_b^0}=d^\gamma_{\Xi_b^0}$. Similarly, in the sextet we obtain
the relations for the qCEDM $d^\gamma_{\Sigma_b^-} + d^\gamma_{\Sigma_b^+} = 2 d^\gamma_{\Sigma_b^0}$, which are also
true for the qEDM, but $d^\gamma_{\Sigma_b^-} \neq d^\gamma_{\Sigma_b^+}$. These differences arise because in order
for the qCEDM to generate an EDM of a baryon, an insertion of the quark charge is required. As such, EDMs
of baryons with a single b quark but different charges differ for the qCEDM. This is not true for
the bottom qEDM as the operator already contains a photon. 

A richer pattern emerges for the four-quark operators as here loop diagrams provide leading contributions.
For instance, for the 4qLR we observe that the tree-level contributions to the triplet and sextet
EDMs have an identical pattern as that of the qCEDM. However, the loop contributions induce differences.
In the triplet, loop contributions lead to a splitting in the EDMs of the neutral baryons and
$d^\gamma_{\Lambda_b^0}\neq d^\gamma_{\Xi_b^0}$, because of the different Goldstone bosons participating in the
loops. We find 
\begin{equation} d^\gamma_{\Lambda_b^0,\mathrm{4qLR}} - d^\gamma_{\Xi_b^0,\mathrm{4qLR}} = \frac{e \mathrm{Re}(V_{ub}) g_2 \nu_3 \nu^{ub} }{16 \pi^2 F_\pi^2} \left( 2 \ln \frac{M_K}{M_\pi} + \frac{2 \Delta {\arccos}\big[\frac{\Delta}{M_K}\big]}{\sqrt{M_K^2-\Delta^2}} - \frac{2 \Delta}{\sqrt{\Delta^2 - M_{\pi}^2 }} \text{ln}\Bigg[ \frac{\Delta}{M_{\pi}}  + \sqrt{\frac{\Delta^2}{M_{\pi}^2} -1 }\Bigg]\right)
\end{equation}
which is nonzero and finite, whereas for the qCEDM this combination vanishes. In the same way, the
degeneracy that is present for the qCEDM for the negatively charged sextet baryons is broken by the
4qLR operator. 
To illustrate this, while for both the qCEDM and the 4qLF we have $d^\gamma_{\Xi^{\prime\,-}_b} -
( d^\gamma_{\Sigma^{-}_b}+d^\gamma_{\Omega^{-}_b})/2=0$, only for the 4qLF $ d^\gamma_{\Sigma^{-}_b}-d^\gamma_{\Omega^{-}_b}
\neq 0$ (and finite). 

Finally, for the 4q operators an even different pattern of EDMs arises depending on the flavor
configuration of the underlying operator. From Eqs.~\eqref{matrix} and \eqref{matrix2} it is clear
that the chiral symmetry properties of $\mu^{ub}$ is identical to the 4qLR operator $\sim \nu^{ub}$.
As such, for $\mu^{ub}$ the same pattern of EDMs emerges as for the 4qLR and these sources cannot be
separated from symmetry arguments alone. Different patterns do emerge for $\mu^{db}$ and $\mu^{sb}$.
For example, the splitting in the triplet is different for $\mu^{db}$ with respect to the 4qLR but
this can probably only be used with additional information on the LECs.

The above considerations indicate that the pattern of EDMs of bottom baryons provide information about
the source of CP violation. If experiments, for instance those proposed in
Refs.~\cite{Fomin:2017ltw,Aiola:2020yam,Bagli:2017foe}, were to see nonzero signals, this information
could be used to pinpoint the underlying mechanism. Much more could be said with nonperturbative
information about the LECs appearing in the Lagrangians. 
\section{How large are the EDMs? }\label{sec:size}
To determine the sizes of the EDMs of bottom baryons as function of the various dimension-six Wilson
coefficients appearing in Eq.~\eqref{pt_vio_op}, estimates of the various LECs appearing in EDM
expressions are necessary. This requires non-perturbative QCD calculations of the associated
matrix elements. While a lot of progress has been made in this direction for EDMs of systems
containing first-generation quarks, see e.g. Ref.~\cite{Shindler:2021bcx} for a recent review, as far
as we know no calculations have been performed for baryons containing heavier valence quarks. In
this work, we estimate the contributions using naive dimensional analysis (NDA), a technique
discussed in detail in Refs.~\cite{Weinberg:1989dx, Manohar:1983md} and used for nucleon EDMs
in Ref.~\cite{deVries:2010ah}. While NDA does not give accurate predictions, it provides a
reasonable order-of-magnitude estimate for meson and single-baryon matrix elements and is
the guiding principle for a systematic power counting in effective field theories. 

The EDMs of the bottom baryons under consideration depend, for each source of CP violation,
on several LECs. The easiest estimate are for the bottom EDM. NDA predicts 
\begin{equation}
c_{1,2} = \mathcal O(d_b) =\mathcal O\left(\frac{m_b}{\Lambda^2}\right)\,, \label{NDA_qEDM}
\end{equation}
which is a rather intuitive result. The bottom quark EDM operator directly induces a bottom baryon
EDM up to order-one factors. The factors could be calculated with non-perturbative methods such as
lattice QCD or estimated using a quark model. For light quarks, for instance, lattice QCD predicts
the neutron EDM to be $d_n = 0.82 d_d - 0.21 d_u$ \cite{Bhattacharya:2015esa} in agreement with NDA
estimates. 

Next, we consider the quark CEDM. In this case we need to estimate the LECs $b_{16}$-$b_{20}$.
NDA predicts
\begin{equation}
 b_{16\mathrm{-}20}= \mathcal O\left(\tilde d_b\frac{F_\pi}{ \Lambda_\chi}\right) =\mathcal O\left(\frac{F_\pi m_b}{\Lambda_\chi\Lambda^2}\right)\,,
\end{equation}
where we used $4 \pi F_\pi \sim \Lambda_\chi$. The loop diagrams only contribute at
next-to-next-to-leading order. It would be interesting to compare these predictions with other
estimates, for instance through QCD sum rules. 

For the four-quark operators we need to estimate both the tree-level LECs as well as the CP-odd
Goldstone boson-baryon interactions. For the $4q$ terms we obtain 
\begin{eqnarray}
\mu_{6\mathrm{-}10}\, \mu^{qb}&=& \mathcal O\left(\mu^{qb} \Lambda_\chi F_\pi^2\right)= \mathcal O \left(\frac{\Lambda_\chi F_\pi^2}{\Lambda^2}\right)\,,\nonumber\\
\mu_{11\mathrm{-}21}\, \mu^{qb}&=& \mathcal O\left(e \mu^{qb} \frac{F_\pi^2}{\Lambda_\chi}\right)= \mathcal O \left(e \frac{F_\pi^2}{\Lambda_\chi \Lambda^2}\right)\,,
\end{eqnarray}
where $q = \{u,\,d,\,s\}$. 
While the Goldstone boson-baryon terms scale as $\sim \Lambda_\chi^1$, and are thus of lower order than
the EDM vertices $\sim \Lambda_\chi^{-1}$, they only contribute to the EDMs at the one-loop level
bringing in a loop factor $e/(4\pi F_\pi)^2\sim e\Lambda_\chi^{-2}$ so that both type of vertices
contribute at the same order. 

Similarly for the 4qLR operator we obtain 
\begin{eqnarray}
\nu_{1\mathrm{-}5}\, \nu^{ub}&=& \mathcal O\left(\nu^{ub} \Lambda_\chi F_\pi^2\right)= \mathcal O \left(\frac{\Lambda_\chi F_\pi^2}{\Lambda^2}\right)\,,\nonumber\\
\nu_{11\mathrm{-}21}\, \nu^{ub}&=& \mathcal O\left(e \nu^{ub} \frac{F_\pi^2}{\Lambda_\chi}\right)= \mathcal O \left(e \frac{F_\pi^2}{\Lambda_\chi \Lambda^2}\right)\,.
\end{eqnarray}

While the NDA estimates are rough they give a reasonable idea of the scale of BSM physics that can
be probed by measuring EDMs of bottom baryons with a given sensitivity.   
For instance, for a BSM physics 
scale\footnote{This scale is comparable to indirect limits obtained from traditional EDM experiments. For example, a b-quark EDM mixes with a b-quark CEDM under the one-loop QED renormalization group. At the b-quark threshold, the latter induces a Weinberg three-gluon operator \cite{Braaten:1990zt} which, in turn, induces a neutron EDM. Based on this procedure Ref.~\cite{Gisbert:2019ftm} quotes an indirect limit $d_b < 1.2 \cdot 10^{-2}$ e cm. Using our parametrization $d_b = m_b/\Lambda^2$ we get $\Lambda >2$ TeV. However, this indirect limit suffers from a large theoretical uncertainty due to the poorly known neutron EDM matrix element of the Weinberg operator \cite{Chien:2015xha,Haisch:2021hcg}. Furthermore, the neutron EDM can get contributions from other sources. We therefore consider $\Lambda=1$ TeV as a reasonable and pragmatic choice.} $\Lambda = 1\,$TeV, and considering only the tree-level expressions we estimate
\begin{equation}
d^\gamma_{B_b} = \{10^{-19}\,, 10^{-20}\,,10^{-21}\,,10^{-24}\}\,e\,\mathrm{cm}\,,
\end{equation}
for the qEDM, qCEDM, 4q, and 4qLR operator, respectively. The smallness of the last term is explained
by the factor of $\text{Re}(V_{ub})$. These estimates involve a sizeable uncertainty. Nevertheless,
they can be used to determine the reach of a potential program to measure the EDMs of bottom-quark
baryons. To get an idea of the uncertainty we used a Monte Carlo (MC) sampling of the LECs that
appear in the EDM expressions. For instance, for the qEDM operator we rescaled the LECs  
\begin{equation}
	c_{1,2} \rightarrow \left( \frac{m_b}{\Lambda^2} \right) \tilde{c}_{1,2} \, , 
\end{equation}
and vary the dimensionless constants $\tilde c_{1,2}$ between $[-3, +3]$. After the MC sampling  we obtain a list of different values for the qEDM contribution from which we compute the mean value and the standard deviation. We use this procedure for all LECs appearing in the EDM expression and obtain the mean values and standard deviations for the various EDMs for each CP-odd source in Tables~\ref{MCEDM-Antitriplet} and~\ref{MCEDM-Sextet}. The MC method is just a tool to determine roughly in what range we can expect an EDM for the various sources at a given scale $\Lambda$.

\begin{table}[t]
	\centering
	\caption{Numerical contributions to the EDMs of the anti-triplet baryons for $\Lambda = 1\,\text{TeV}$.
          The results are given in $10^{-19}\,e\,\text{cm}$, $10^{-20}\,e\,\text{cm}$, $10^{-21}\,e\,\text{cm}$,
          and $10^{-24}\,e\,\text{cm}$ for the qEDM-, qCEDM-, 4q-, and 4qLR-operator, respectively.}
	\begin{tabular}{lccc}
		\toprule
		{Contribution}   &  {$\Lambda_{b}^0$}    & {$\Xi_{b}^0$} &   {$\Xi_{b}^-$}   \\
		\midrule  
		{qEDM}  & $ -0.24 \pm 5.7 $ & $ -0.24 \pm 5.7$ & $ -0.24 \pm 5.7 $ \\
		{qCEDM}  & $+0.18 \pm 4.6 $ & $+0.18 \pm 4.6 $ & $+0.40 \pm 6.5$ \\
		{4q}  & $-0.070 \pm 2.4$ & $-0.020 \pm 2.5$ & $+0.040 \pm 3.2$ \\
		{4qLR}  & $+0.15 \pm 9.4$ & $+0.58 \pm 9.6$ & $-0.11 \pm 10.8$ \\
		\bottomrule
	\end{tabular}
	\label{MCEDM-Antitriplet}
\end{table}

\begin{table}[t]
	\centering
	\caption{Numerical contributions to the EDMs of the sextet baryons for $\Lambda = 1\, \text{TeV}$.
          The results are given in $10^{-19}\,e\,\text{cm}$, $10^{-20}\,e\,\text{cm}$, $10^{-21}\,e\,\text{cm}$,
          and $10^{-24}\,e\,\text{cm}$ for the qEDM-, qCEDM-, 4q-, and 4qLR-operator, respectively.}
	\begin{tabular}{lcccccc}
		\toprule
		{Contribution}   &  {$\Sigma_{b}^+$}  & {$\Sigma_{b}^0$}  & {$\Sigma_{b}^-$} & {$\Xi_{b}^{'0}$} & {$\Xi_{b}^{'-}$} & {$\Omega_{b}^-$}   \\
		\midrule  
		{qEDM}  & $-0.16 \pm 2.8$ & $-0.16 \pm 2.8$ & $-0.16 \pm 2.8$ & $-0.1 \pm 2.8$ & $-0.16 \pm 2.8$ & $-0.16 \pm 2.8$ \\
		{qCEDM}  & $+0.10 \pm 3.3$ & $+0.04 \pm 2.2$ & $+0.070 \pm 3.3$ & $+0.040 \pm 2.2$ & $+0.070 \pm 3.3$ & $+0.070 \pm 3.3$ \\
		{4q}  & $-0.050 \pm 2.1$ & $+0.070 \pm 1.2$ & $+0.040 \pm 2.1$ & $+0.020 \pm 1.3$ & $+0.050 \pm 1.6$ & $-0.060 \pm 2.0$ \\
		{4qLR}  & $-0.23 \pm 7.9$ & $-0.010 \pm 4.9$ & $+0.21 \pm 5.7$ & $+0.050 \pm 4.8$ & $+0.21 \pm 5.6$ & $+0.35 \pm 5.3$ \\
		\bottomrule
	\end{tabular}
	\label{MCEDM-Sextet}
\end{table}

In Table~\ref{MCEDM-Antitriplet} we collect the different contributions to the EDMs of the anti-triplet states.
For each source, we get, unsurprisingly, results that vary around zero with a spread given by the NDA
estimates. There is roughly an order-of-magnitude uncertainty. As expected, the qEDM dominates, whereas
the 4qLR-term gives the smallest contribution. The standard deviations for all contributions are
relatively large, which is explained by the wide range of the dimensionless constants which was
used in the MC sampling. The same observations can also be drawn from Table~\ref{MCEDM-Sextet}.
In the case that all four dimension-six operators contribute at the same BSM scale $\Lambda$, we
can take a look at the resulting size of the EDM by adding up the single contributions. Taking
the $\Omega_{b}^{-}$ baryon as an example, the total EDM would be 
\begin{equation}
d^{\gamma}_{\Omega_{b}^{-}} = (-0.15 \pm 3.2) \times 10^{-19} \left( \text{TeV} / \Lambda \right)^2 \,e\,\text{cm} \, .
\end{equation}
This value is of course not to be understood as a clear prediction, but as an estimate for the
range where the EDM of the $\Omega_{b}^{-}$ baryon can be found. The experiment would involve the positively charged anti-baryons (e.g. $\Omega_{b}^{+}$) whose EDMs are the same as the corresponding baryons by CPT. The errors here reflect the uncertainty on the hadronic theory and, while the error band includes zero, nothing would indicate a vanishing matrix element. 
For the qEDM and qCEDM operators, indirect limits have been set from the EDM of the neutron and
diamagnetic atoms \cite{Braaten:1990zt,Chien:2015xha,Gisbert:2019ftm,Haisch:2021hcg}. We do not
compare these limits here in detail as the indirect limits are plagued by sizeable uncertainties
as well (mainly from matrix elements connecting the three-gluon Weinberg operator to the neutron EDM)
and assume that there are not other contributions to the neutron EDM (for instance from EDMs or CEDMs
of light quarks). Our main goal here is to assess the reach of a potential experimental program to
measure the EDMs of bottom-quark baryons.
\clearpage
\section{Conclusion}
\label{sec:con}
Electric dipole moment experiments provide one of the most sensitive searches for BSM physics.
Most focus has been on EDMs of stable systems consisting of first-generation quarks, but 
it has been proposed to look for EDMs of baryons containing heavier quarks as well
\cite{Fomin:2017ltw,Aiola:2020yam,Bagli:2017foe}. 
Such systems are sensitive to CP-odd operators involving second- and third-generation quarks and
complement existing searches.  However, essentially no theoretical calculations have been performed
to guide this developing experimental program. 

In this paper, we have analyzed the EDMs of spin-$1/2$ baryons containing a single bottom quark.
Our starting point has been operators of dimension-six in the SMEFT Lagrangian that violate CP and
contain a $\bar b \Gamma b$ bilinear (where $\Gamma$ denotes a Lorentz structure). 
We considered a hypothetical bottom quark EDM and chromo-EDM, and several four-quark operators
mixing bottom quarks with lighter quarks. We used chiral perturbation theory to construct the resulting
CP-violating hadronic interactions between spin-$1/2$ single-bottom baryons, Goldstone bosons, and
photons, and calculated the EDMs up to the first non-vanishing order for each source of CP violation. 

Our results indicate that different sources of CP violation lead to a different pattern of EDMs due
to the chiral- and isospin-symmetry properties of the underlying sources. In principle, this would
allow for the identification of the dominant source of CP violation based on the relative sizes of EDMs
of triplet and sextet bottom-quark baryons. The absolute sizes of the EDMs, however, are very uncertain as
very little is known about the magnitudes of the low-energy constants appearing in the CP-odd chiral
Lagrangian. We made estimates using naive dimensional analysis and found that for BSM scales of $1$~TeV,
we can expect EDMs in the range of $10^{-19}-10^{-24}$ e~cm depending strongly on the dimension-six
operators under consideration. All EDMs scale as $\Lambda^{-2}$ so the sizes of the EDMs can easily be
obtained for other BSM scales. If the experimental program picks up steam and EDMs of these systems are
targeted it would be good to calculate the LECs with non-perturbative techniques to get more reliable
estimates. The techniques developed in this work can be readily extended to calculate EDMs of charmed
baryons and work along these lines is in progress. 

\section*{Acknowledgements}

Partial financial support from the DFG (Project number 196253076 - TRR 110)
and the NSFC (Grant No. 11621131001) through the funds provided
to the Sino-German CRC 110 ``Symmetries and the Emergence of
Structure in QCD",  by the Chinese Academy of Sciences (CAS) through a President's
International Fellowship Initiative (PIFI) (Grant No. 2018DM0034), by the VolkswagenStiftung
(Grant No. 93562), and by the EU Horizon 2020 research and innovation programme, STRONG-2020 project
(Grant No. 824093) is acknowledged.
\clearpage
%
%

\begin{appendix}
\section{Form Factors}
\label{sec:AppendixA}
The full expression for the neutral and charged $b$-baryon form factors up to the order $\mathcal{O}(\delta^2)$ with the tree-level results are 
\begin{equation}
\vspace{70pt}
	\begin{aligned} 
	D^{\gamma}_{\Lambda_b^0}(q^2)= ~& 4c_1 -4e\Big(  b_{19} -\mu_{11}(\mu^{ub}-\mu^{db})+\mu_{14}(\mu^{ub}+\mu^{db}) -2\mu_{20}(\mu^{ub}-\mu^{db}-\mu^{sb}) \\
	& -\text{Re}(V_{ub})(\nu_{11}-\nu_{14}+2\nu_{20}) \nu^{ub} \Big) \\
	      & + \frac{e g_2}{4 F_{\pi}^2}\Big( \text{Re}(V_{ub}) \nu_3 \nu^{ub} +\mu_8 ( \mu^{ub}+\mu^{sb}) \Big) \int_0^1 dx \, \frac{1}{\tilde{M}_K}\frac{\partial}{\partial \tilde{M}_K}J_1( -\Delta, \tilde{M}_K) , 
	      \\
	D^{\gamma}_{\Xi_b^0}(q^2)= ~& 4 c_1 -4e\Big(  b_{19}-\mu_{11}(\mu^{ub}-\mu^{sb})+\mu_{14}(\mu^{ub}+\mu^{sb})-2\mu_{20}(\mu^{ub}-\mu^{db}-\mu^{sb})  \\
		& -\text{Re}(V_{ub})(\nu_{11}-\nu_{14} +2\nu_{20}) \nu^{ub} \Big) \\
		 & + \frac{e g_2}{4 F_{\pi}^2}\Big( \text{Re}(V_{ub}) \nu_3 \nu^{ub} +\mu_8 ( \mu^{ub}+\mu^{db}) \Big) \int_0^1 dx \, \frac{1}{\tilde{M}_{\pi}}\frac{\partial}{\partial \tilde{M}_{\pi}}J_1(-\Delta, \tilde{M}_{\pi}) ,  \\
	D^{\gamma}_{\Xi_b^-}(q^2)=  ~& 4 c_1 -4e\Big(  b_{16} + b_{19} + (\mu_{11}+\mu_{14})(\mu^{db}+\mu^{sb}) + 2 \mu_{17}(\mu^{ub}+\mu^{db}+\mu^{sb}) \\ 
	& - 2\mu_{20}(\mu^{ub}-\mu^{db}-\mu^{sb})+ 2 \text{Re}(V_{ub})(\nu_{17} - \nu_{20}) \nu^{ub} \Big ) \\
	& - \frac{e g_2}{4 \pi^2 F_{\pi}^2}\Big( \text{Re}(V_{ub}) \nu_3 \nu^{ub} + \mu_8 ( \mu^{ub}+\mu^{sb}) \Big) \int_0^1 dx \, \frac{1}{\tilde{M}_K}\frac{\partial}{\partial \tilde{M}_K}J_1( -\Delta, \tilde{M}_K) \\
	& - \frac{e g_2}{4 \pi^2 F_{\pi}^2}\Big( \text{Re}(V_{ub}) \nu_3 \nu^{ub} +\mu_8 ( \mu^{ub}+\mu^{db}) \Big) \int_0^1 dx \, \frac{1}{\tilde{M}_{\pi}}\frac{\partial}{\partial \tilde{M}_{\pi}}J_1( -\Delta, \tilde{M}_{\pi}) ,   
	\\
	D^{\gamma}_{\Sigma_b^+}(q^2)=  ~& 2c_2 + 2e\Big( b_{17} - b_{20} +2 (\mu_{12}-\mu_{15}) \mu^{ub}+2\mu_{18}(\mu^{ub}+\mu^{db}+\mu^{sb}) \\
	&+ 2 \mu_{21} (\mu^{ub}-\mu^{db}-\mu^{sb}) +2 \text{Re}(V_{ub})(\nu_{12}-\nu_{15}+\nu_{18}+\nu_{21}) \nu^{ub} \Big) \\
	& + \frac{e g_1}{4 \pi^2 F_{\pi}^2}\Big( \text{Re}(V_{ub}) \nu_2 \nu^{ub} +\mu_7 ( \mu^{ub}+\mu^{sb}) \Big)  \int_0^1 dx \, \frac{1}{\tilde{M}_K}\frac{\partial}{\partial \tilde{M}_K}J_1( -\Delta, \tilde{M}_K) \\
	& + \frac{e g_1}{4 \pi^2 F_{\pi}^2}\Big( \text{Re}(V_{ub}) \nu_2 \nu^{ub} +\mu_7 ( \mu^{ub}+\mu^{db}) \Big)  \int_0^1 dx \, \frac{1}{\tilde{M}_{\pi}}\frac{\partial}{\partial \tilde{M}_{\pi}}J_1( -\Delta, \tilde{M}_{\pi}) \\
	& + \frac{e g_2}{2 \pi^2 F_{\pi}^2} \Big( \text{Re}(V_{ub}) \nu_3 \nu^{ub} +\mu_8 ( \mu^{ub}+\mu^{sb}) \Big)  \int_0^1 dx \, \frac{1}{\tilde{M}_K}\frac{\partial}{\partial \tilde{M}_K}J_1( 0, \tilde{M}_K) \\
	& + \frac{e g_2}{2 \pi^2 F_{\pi}^2} \Big( \text{Re}(V_{ub}) \nu_3 \nu^{ub} +\mu_8 ( \mu^{ub}+\mu^{db}) \Big) \int_0^1 dx \, \frac{1}{\tilde{M}_{\pi}}\frac{\partial}{\partial \tilde{M}_{\pi}}J_1( 0, \tilde{M}_{\pi})  , \\
	D^{\gamma}_{\Sigma_b^0}(q^2)=  ~&  2 c_2 -2e\Big( b_{20} - \mu_{12} (\mu^{ub}-\mu^{db})+\mu_{15}(\mu^{ub}+\mu^{db})- 2 \mu_{21} (\mu^{ub}-\mu^{db}-\mu^{sb}) \\
	& - \text{Re}(V_{ub})(\nu_{12}-\nu_{15} +2 \nu_{21}) \nu^{ub} \Big) \\
	&+ \frac{e g_1}{8 F_{\pi}^2}\Big( \text{Re}(V_{ub}) \nu_2 \nu^{ub} +\mu_7 ( \mu^{ub}+\mu^{sb}) \Big) \int_0^1 dx \, \frac{1}{\tilde{M}_K}\frac{\partial}{\partial \tilde{M}_K}J_1( -\Delta, \tilde{M}_K) \\
	& + \frac{e g_2}{4 F_{\pi}^2}\Big( \text{Re}(V_{ub}) \nu_3 \nu^{ub} +\mu_8 ( \mu^{ub}+\mu^{sb})  \Big) \int_0^1 dx \, \frac{1}{\tilde{M}_K}\frac{\partial}{\partial \tilde{M}_K}J_1(0, \tilde{M}_K) , 
	\nonumber
	\end{aligned}
	\label{all-edm-form}
\end{equation} 
\begin{equation}
	\begin{aligned}
	D^{\gamma}_{\Sigma_b^-}(q^2)=  ~& 2 c_2 -2e\Big(  b_{17} + b_{20} +2 (\mu_{12}+\mu_{15})\mu^{db}+ 2 \mu_{18}(\mu^{ub}+\mu^{db}+\mu^{sb})\\
		& -2 \mu_{21} (\mu^{ub}-\mu^{db}-\mu^{sb}) + 2 \text{Re}(V_{ub})(\nu_{18} -\nu_{21} ) \nu^{ub} \Big) \\ 
		& - \frac{e g_1}{4 \pi^2 F_{\pi}^2}\Big( \text{Re}(V_{ub}) \nu_2 \nu^{ub} +\mu_7 ( \mu^{ub}+\mu^{db}) \Big)  \int_0^1 dx \, \frac{1}{\tilde{M}_{\pi}}\frac{\partial}{\partial \tilde{M}_{\pi}}J_1( -\Delta, \tilde{M}_{\pi}) \\
		& - \frac{e g_2}{2 \pi^2 F_{\pi}^2} \Big( \text{Re}(V_{ub}) \nu_3 \nu^{ub} +\mu_8 ( \mu^{ub}+\mu^{db}) \Big)  \int_0^1 dx \, \frac{1}{\tilde{M}_{\pi}}\frac{\partial}{\partial \tilde{M}_{\pi}}J_1( 0, \tilde{M}_{\pi}) ,  \\
		D^{\gamma}_{\Xi_b^{'0}}(q^2)=  ~&  2 c_2 -2e\Big( b_{20} - \mu_{12}(\mu^{ub}-\mu^{sb})+\mu_{15}(\mu^{ub}+\mu^{sb}) - 2 \mu_{21} ( \mu^{ub}-\mu^{db}-\mu^{sb} ) \\
		& - \text{Re}(V_{ub})(\nu_{12}-\nu_{15} + 2 \nu_{21}) \nu^{ub} \Big) \\
		& + \frac{e g_1}{8 F_{\pi}^2}\Big( \text{Re}(V_{ub}) \nu_2 \nu^{ub} +\mu_7 ( \mu^{ub}+\mu^{db}) \Big) \int_0^1 dx \, \frac{1}{\tilde{M}_{\pi}}\frac{\partial}{\partial \tilde{M}_{\pi}}J_1( -\Delta, \tilde{M}_{\pi}) \\
						& + \frac{e g_2}{4 F_{\pi}^2}\Big( \text{Re}(V_{ub}) \nu_3 \nu^{ub} +\mu_8 ( \mu^{ub}+\mu^{db}) \Big) \int_0^1 dx \, \frac{1}{\tilde{M}_\pi}\frac{\partial}{\partial \tilde{M}_\pi}J_1(0, \tilde{M}_\pi) ,  
		  \\
	  D^{\gamma}_{\Xi_b^{'-}}(q^2)=  ~& 2 c_2 -2e\Big( b_{17} + b_{20} + (\mu_{12}+\mu_{15})(\mu^{db}+\mu^{sb})+2 \mu_{18}(\mu^{ub}+\mu^{db}+\mu^{sb}) \\
		& - 2 \mu_{21} (\mu^{ub}-\mu^{db}-\mu^{sb}) +2 \text{Re}(V_{ub})(\nu_{18} - \nu_{21}) \nu^{ub} \Big) \\
		& - \frac{e g_1}{8 \pi^2 F_{\pi}^2}\Big( \text{Re}(V_{ub}) \nu_2 \nu^{ub} +\mu_7 ( \mu^{ub}+\mu^{sb}) \Big) \int_0^1 dx \, \frac{1}{\tilde{M}_K}\frac{\partial}{\partial \tilde{M}_K}J_1( -\Delta, \tilde{M}_K) \\
		& - \frac{e g_1}{8 \pi^2 F_{\pi}^2}\Big( \text{Re}(V_{ub}) \nu_2 \nu^{ub} +\mu_7 ( \mu^{ub}+\mu^{db}) \Big) \int_0^1 dx \, \frac{1}{\tilde{M}_{\pi}}\frac{\partial}{\partial \tilde{M}_{\pi}}J_1( -\Delta, \tilde{M}_{\pi}) \\
		 & - \frac{e g_2}{4 \pi^2 F_{\pi}^2} \Big( \text{Re}(V_{ub}) \nu_3 \nu^{ub} +\mu_8 ( \mu^{ub}+\mu^{sb}) \Big) \int_0^1 dx \, \frac{1}{\tilde{M}_K}\frac{\partial}{\partial \tilde{M}_K}J_1( 0, \tilde{M}_K) \\
		& - \frac{e g_2}{4 \pi^2 F_{\pi}^2} \Big( \text{Re}(V_{ub}) \nu_3 \nu^{ub}+\mu_8 ( \mu^{ub}+\mu^{db}) \Big) \int_0^1 dx \, \frac{1}{\tilde{M}_{\pi}}\frac{\partial}{\partial \tilde{M}_{\pi}}J_1( 0, \tilde{M}_{\pi}) ,
		\\
		D^{\gamma}_{\Omega_b^-}(q^2)=  ~& 2 c_2 -2e\Big( b_{17} + b_{20} +2 (\mu_{12}+\mu_{15})\mu^{sb}+ 2\mu_{18}(\mu^{ub}+\mu^{db}+\mu^{sb}) \\
		& - 2 \mu_{21} (\mu^{ub}-\mu^{db}-\mu^{sb})  + 2 \text{Re}(V_{ub})(\nu_{18} - \nu_{21} )\nu^{ub} \Big) \\
		& - \frac{e g_1}{4 \pi^2 F_{\pi}^2}\Big( \text{Re}(V_{ub}) \nu_2 \nu^{ub} +\mu_7 ( \mu^{ub}+\mu^{sb}) \Big)  \int_0^1 dx \, \frac{1}{\tilde{M}_K}\frac{\partial}{\partial \tilde{M}_K}J_1( -\Delta, \tilde{M}_K) \\
		& - \frac{e g_2}{2 \pi^2 F_{\pi}^2} \Big( \text{Re}(V_{ub}) \nu_3 \nu^{ub} +\mu_8 ( \mu^{ub}+\mu^{sb}) \Big) \int_0^1 dx \, \frac{1}{\tilde{M}_K}\frac{\partial}{\partial \tilde{M}_K}J_1( 0, \tilde{M}_K) .
	\end{aligned}
\end{equation}

\section{EDMs with NDA Estimates}
\label{sec:AppendixB}
Replacing the unknown LECs in the equations for the neutral and charged $b$-baryon EDMs with the NDA estimate leads to the following expressions 
\begin{equation}
\begin{aligned} 
d^{\gamma}_{\Lambda_b^0}= ~& 4 \Bigg( \frac{m_b}{\Lambda^2} \Bigg) \tilde{c}_1 - 4e \Bigg( \frac{m_b}{4 \pi \Lambda^2}\Bigg) \tilde{b}_{19} + 4e \Bigg( \frac{F_{\pi}}{4 \pi \Lambda^2}\Bigg) \big[ \tilde{\mu}_{11} - \tilde{\mu}_{14} + 2 \tilde{\mu}_{20} + \text{Re}(V_{ub}) ( \tilde{\nu}_{11} - \tilde{\nu}_{14} + 2 \tilde{\nu}_{20} ) \big] \\
 		 & + \frac{e g_2}{32 \pi^2} \frac{\Lambda_\chi}{\Lambda^2} \Big(  \text{Re}(V_{ub}) \tilde{\nu}_{3} + \tilde{\mu}_{8} \Big) 
		\Bigg( 1 + 2 \text{ln}\Bigg[\frac{M_K}{\lambda}\Bigg] + \frac{2 \Delta \text{Arccos}\big[\frac{\Delta}{M_K}\big]}{\sqrt{M_K^2-\Delta^2}} \Bigg), \\
d^{\gamma}_{\Xi_b^0}= ~& 4 \Bigg( \frac{m_b}{\Lambda^2} \Bigg) \tilde{c}_1 - 4e \Bigg( 		\frac{m_b}{4 \pi \Lambda^2}\Bigg) \tilde{b}_{19} + 4e \Bigg( \frac{F_{\pi}}{4 \pi \Lambda^2}\Bigg) \big[ \tilde{\mu}_{11} - \tilde{\mu}_{14} + 2 \tilde{\mu}_{20} + \text{Re}(V_{ub}) ( \tilde{\nu}_{11} - \tilde{\nu}_{14} + 2 \tilde{\nu}_{20} ) \big] \\
		& + \frac{e g_2}{32 \pi^2} \frac{\Lambda_\chi}{\Lambda^2} \Big(  \text{Re}(V_{ub}) \tilde{\nu}_{3} + \tilde{\mu}_{8} \Big) \Bigg( 1 + 2 \text{ln}\Bigg[\frac{M_{\pi}}{\lambda}\Bigg] + \frac{2 \Delta}{\sqrt{\Delta^2-M_{\pi}^2}} \text{ln}\Bigg[ \frac{\Delta}{M_{\pi}}  + \sqrt{\frac{\Delta^2}{M_{\pi}^2} -1 } \Bigg] \Bigg), \\                   
d^{\gamma}_{\Xi_b^-} = ~& 4 \Bigg( \frac{m_b}{\Lambda^2} \Bigg) \tilde{c}_1 - 4e \Bigg( 		\frac{m_b}{4 \pi \Lambda^2}\Bigg) (\tilde{b}_{16} + \tilde{b}_{19}) \\
		& - 4e \Bigg( \frac{F_{\pi}}{4 \pi \Lambda^2}\Bigg) \big[ \tilde{\mu}_{11} + \tilde{\mu}_{14} +2 \tilde{\mu}_{17} - 2 \tilde{\mu}_{20} + 2 \text{Re}(V_{ub}) ( \tilde{\nu}_{17} - \tilde{\nu}_{20} ) \big]  - \frac{e g_2}{32 \pi^2} \frac{\Lambda_\chi}{\Lambda^2} \Big(  \text{Re}(V_{ub}) \tilde{\nu}_{3} + \tilde{\mu}_{8} \Big) \\
		& \times \Bigg( 2 + 2 \text{ln}\Bigg[\frac{M_K}{\lambda}\Bigg] + \frac{2 \Delta \text{Arccos}\big[\frac{\Delta}{M_K}\big]}{\sqrt{M_K^2-\Delta^2}} + 2 \text{ln}\Bigg[\frac{M_{\pi}}{\lambda}\Bigg] + \frac{2 \Delta}{\sqrt{\Delta^2-M_{\pi}^2}} \text{ln}\Bigg[ \frac{\Delta}{M_{\pi}}  + \sqrt{\frac{\Delta^2}{M_{\pi}^2} -1 } \Bigg]  \Bigg), \\ 
d^{\gamma}_{\Sigma_b^+} = ~& 2 \Bigg( \frac{m_b}{\Lambda^2} \Bigg) \tilde{c}_2 + 2e \Bigg( \frac{m_b}{4 \pi \Lambda^2}\Bigg) (\tilde{b}_{17} - \tilde{b}_{20}) \\
		& + 4e \Bigg( \frac{F_{\pi}}{4 \pi \Lambda^2}\Bigg) \big[ \tilde{\mu}_{12} - \tilde{\mu}_{15} + \tilde{\mu}_{18} + \tilde{\mu}_{21} + \text{Re}(V_{ub}) ( \tilde{\nu}_{12} - \tilde{\nu}_{15} + \tilde{\nu}_{18} + \tilde{\nu}_{21} ) \big] \\
		& + \frac{e g_2}{16 \pi^2} \frac{\Lambda_\chi}{\Lambda^2} \Big(  \text{Re}(V_{ub}) \tilde{\nu}_{3} + \tilde{\mu}_{8} \Big) \Bigg( 2 + 2 \text{ln}\Bigg[\frac{M_K M_{\pi} }{\lambda^2}\Bigg] \Bigg) + \frac{e g_1}{32 \pi^2} \frac{\Lambda_\chi}{\Lambda^2} \Big(  \text{Re}(V_{ub}) \tilde{\nu}_{2} + \tilde{\mu}_{7} \Big) \\
		& \times \Bigg( 2 + 2 \text{ln}\Bigg[\frac{M_K}{\lambda}\Bigg] + \frac{2 \Delta \text{Arccos}\big[\frac{\Delta}{M_K}\big]}{\sqrt{M_K^2-\Delta^2}} + 2 \text{ln}\Bigg[\frac{M_{\pi}}{\lambda}\Bigg] + \frac{2 \Delta}{\sqrt{\Delta^2-M_{\pi}^2}} \text{ln}\Bigg[ \frac{\Delta}{M_{\pi}}  + \sqrt{\frac{\Delta^2}{M_{\pi}^2} -1 } \Bigg]  \Bigg), \\
d^{\gamma}_{\Sigma_b^0}=  ~&  2 \Bigg( \frac{m_b}{\Lambda^2} \Bigg) \tilde{c}_2 - 2e \Bigg( \frac{m_b}{4 \pi \Lambda^2}\Bigg)  \tilde{b}_{20} + 2e \Bigg( \frac{F_{\pi}}{4 \pi \Lambda^2}\Bigg) \big[ \tilde{\mu}_{12} - \tilde{\mu}_{15} + 2 \tilde{\mu}_{21} + \text{Re}(V_{ub}) ( \tilde{\nu}_{12} - \tilde{\nu}_{15} + 2 \tilde{\nu}_{21} ) \big] \\ 
& + \frac{e g_1}{64 \pi^2} \frac{\Lambda_\chi}{\Lambda^2} \Big(  \text{Re}(V_{ub}) \tilde{\nu}_{2} + \tilde{\mu}_{7} \Big) 
\Bigg( 1 + 2 \text{ln}\Bigg[\frac{M_K}{\lambda}\Bigg] + \frac{2 \Delta \text{Arccos}\big[\frac{\Delta}{M_K}\big]}{\sqrt{M_K^2-\Delta^2}} \Bigg) \\
& + \frac{e g_2}{32 \pi^2}\frac{\Lambda_\chi}{\Lambda^2} \Big( \text{Re}(V_{ub}) \tilde{\nu}_{3} + \tilde{\mu}_{8} \Big) \Bigg(1 + 2 \text{ln} \Bigg[\frac{M_K}{\lambda}\Bigg]\Bigg), \\
d^{\gamma}_{\Sigma_b^-} = ~& 2 \Bigg( \frac{m_b}{\Lambda^2} \Bigg) \tilde{c}_2 - 2e \Bigg( \frac{m_b}{4 \pi \Lambda^2}\Bigg) (\tilde{b}_{17} + \tilde{b}_{20}) - 4e \Bigg( \frac{F_{\pi}}{4 \pi \Lambda^2}\Bigg) \big[ \tilde{\mu}_{12} + \tilde{\mu}_{15} + \tilde{\mu}_{18} - \tilde{\mu}_{21} + \text{Re}(V_{ub}) ( \tilde{\nu}_{18} - \tilde{\nu}_{21} ) \big] \\ 
& - \frac{e g_1}{32 \pi^2} \frac{\Lambda_\chi}{\Lambda^2} \Big(  \text{Re}(V_{ub}) \tilde{\nu}_{2} + \tilde{\mu}_{7} \Big) \Bigg( 1 + 2 \text{ln}\Bigg[\frac{M_{\pi}}{\lambda}\Bigg] + \frac{2 \Delta}{\sqrt{\Delta^2-M_{\pi}^2}} \text{ln}\Bigg[ \frac{\Delta}{M_{\pi}} + \sqrt{\frac{\Delta^2}{M_{\pi}^2} -1 } \Bigg] \Bigg) \\
& - \frac{e g_2}{16 \pi^2} \frac{\Lambda_\chi}{\Lambda^2} \Big(  \text{Re}(V_{ub}) \tilde{\nu}_{3} + \tilde{\mu}_{8} \Big) \Bigg( 1 + 2 \text{ln}\Bigg[\frac{ M_{\pi} }{\lambda}\Bigg] \Bigg)  ,  
	\nonumber
\label{all-edm-nda}
\end{aligned} 
\end{equation}
\begin{equation}
	\begin{aligned}
	d^{\gamma}_{\Xi_b^{'0}}=  ~&  2 \Bigg( \frac{m_b}{\Lambda^2} \Bigg) \tilde{c}_2 - 2e \Bigg( \frac{m_b}{4 \pi \Lambda^2}\Bigg)  \tilde{b}_{20} + 2e \Bigg( \frac{F_{\pi}}{4 \pi \Lambda^2}\Bigg) \big[ \tilde{\mu}_{12} - \tilde{\mu}_{15} + 2 \tilde{\mu}_{21} + \text{Re}(V_{ub}) ( \tilde{\nu}_{12} - \tilde{\nu}_{15} + 2 \tilde{\nu}_{21} ) \big] \\ 
			& + \frac{e g_1}{64 \pi^2} \frac{\Lambda_\chi}{\Lambda^2} \Big(  \text{Re}(V_{ub}) \tilde{\nu}_{2} + \tilde{\mu}_{7} \Big) \Bigg( 1 + 2 \text{ln}\Bigg[\frac{M_{\pi}}{\lambda}\Bigg] + \frac{2 \Delta}{\sqrt{\Delta^2-M_{\pi}^2}} \text{ln}\Bigg[ \frac{\Delta}{M_{\pi}} + \sqrt{\frac{\Delta^2}{M_{\pi}^2} -1 } \Bigg] \Bigg) \\ 
			&+ \frac{e g_2}{32 \pi^2} \frac{\Lambda_\chi}{\Lambda^2} \Big(  \text{Re}(V_{ub}) \tilde{\nu}_{3} + \tilde{\mu}_{8} \Big) \Bigg(1+ 2 \text{ln} \Bigg[\frac{M_\pi}{\lambda}\Bigg]\Bigg) , \\
		d^{\gamma}_{\Xi_b^{'-}} = ~& 2 \Bigg( \frac{m_b}{\Lambda^2} \Bigg) \tilde{c}_2 - 2e \Bigg( \frac{m_b}{4 \pi \Lambda^2}\Bigg) (\tilde{b}_{17} + \tilde{b}_{20}) \\ 
				& - 2e \Bigg( \frac{F_{\pi}}{4 \pi \Lambda^2}\Bigg) \big[ \tilde{\mu}_{12} + \tilde{\mu}_{15} + 2 \tilde{\mu}_{18} - 2 \tilde{\mu}_{21} + 2 \text{Re}(V_{ub}) ( \tilde{\nu}_{18} - \tilde{\nu}_{21} ) \big] \\
				& - \frac{e g_2}{32 \pi^2} \frac{\Lambda_\chi}{\Lambda^2} \Big(  \text{Re}(V_{ub}) \tilde{\nu}_{3} + \tilde{\mu}_{8} \Big) \Bigg( 2 + 2 \text{ln}\Bigg[\frac{M_K M_{\pi} }{\lambda^2}\Bigg] \Bigg) - \frac{e g_1}{64 \pi^2} \frac{\Lambda_\chi}{\Lambda^2} \Big(  \text{Re}(V_{ub}) \tilde{\nu}_{2} + \tilde{\mu}_{7} \Big) \\
				& \times \Bigg( 2 + 2 \text{ln}\Bigg[\frac{M_K}{\lambda}\Bigg] + \frac{2 \Delta \text{Arccos}\big[\frac{\Delta}{M_K}\big]}{\sqrt{M_K^2-\Delta^2}} + 2 \text{ln}\Bigg[\frac{M_{\pi}}{\lambda}\Bigg] + \frac{2 \Delta}{\sqrt{\Delta^2-M_{\pi}^2}} \text{ln}\Bigg[ \frac{\Delta}{M_{\pi}}  + \sqrt{\frac{\Delta^2}{M_{\pi}^2} -1 } \Bigg] \Bigg), \\
		d^{\gamma}_{\Omega_b^-} = ~& 2 \Bigg( \frac{m_b}{\Lambda^2} \Bigg) \tilde{c}_2 - 2e \Bigg( \frac{m_b}{4 \pi \Lambda^2}\Bigg) (\tilde{b}_{17} + \tilde{b}_{20}) - 4e \Bigg( \frac{F_{\pi}}{4 \pi \Lambda^2}\Bigg) \big[ \tilde{\mu}_{12} + \tilde{\mu}_{15} + \tilde{\mu}_{18} - \tilde{\mu}_{21} + \text{Re}(V_{ub}) ( \tilde{\nu}_{18} - \tilde{\nu}_{21} ) \big]  \\
				& - \frac{e g_1}{32 \pi^2} \frac{\Lambda_\chi}{\Lambda^2} \Big( \text{Re}(V_{ub}) \tilde{\nu}_{2} + \tilde{\mu}_{7} \Big) \Bigg( 1 + 2 \text{ln}\Bigg[\frac{M_K}{\lambda}\Bigg] + \frac{2 \Delta \text{Arccos}\big[\frac{\Delta}{M_K}\big]}{\sqrt{M_K^2-\Delta^2}} \Bigg) \\
				& - \frac{e g_2}{16 \pi^2} \frac{\Lambda_\chi}{\Lambda^2} \Big(  \text{Re}(V_{ub}) \tilde{\nu}_{3} + \tilde{\mu}_{8} \Big) \Bigg( 1 + 2 \text{ln}\Bigg[\frac{ M_K }{\lambda}\Bigg] \Bigg) , 
	\end{aligned}
\end{equation}
where $\tilde{c}_i$, $\tilde{b}_i$, $\tilde{\mu}_i$, and $\tilde{\nu}_i$ are dimensionless constants which are varied from $-3$ to $+3$ in the MC sampling. The estimation $4 \pi F_\pi \sim \Lambda_\chi$ has been used at certain steps. 
\clearpage
\section{Loop Functions}
\label{sec:AppendixC}

In this appendix we give the loop functions in the heavy baryon formulation \cite{Bernard:1995dp}
which appear in the calculation of the diagrams in Fig.~\ref{fig:diag} 
\begin{equation}
\begin{aligned}
    \Delta_{M}=~& 2 M^2 \, \Bigg[ L + \frac{1}{16 \pi^2} \text{ln}\Bigg(\frac{M}{\lambda}\Bigg)\Bigg]+\mathcal{O}(n-4), \\
   \frac{1}{i} \bigintssss \frac{d^n k}{(2 \pi)^n} & \frac{1}{M^2-k^2} = ~\Delta_{M}=M^{n-2}\,(4 \pi)^{-n/2}\, \Gamma \Big(1-\frac{n}{2}\Big),
\end{aligned}
\end{equation}	
\begin{equation}
	 \frac{1}{i} \bigintssss \frac{d^n k}{(2 \pi)^n} \frac{\{1,k_{\mu}, k_{\mu} k_{ \nu}\}}{[v \cdot k-w][M^2-k^2]}=\Big\{J_0(w), v_{\mu} J_1(w), g_{\mu \nu} J_2(w)+v_{\mu} v_{\nu} J_3(w)\Big\} ,
\end{equation}
\begin{equation}
	\frac{1}{i} \bigintssss \frac{d^n k}{(2 \pi)^n} \frac{\{1,k_{\mu}, k_{\mu} k_{ \nu}\}}{[v \cdot k-w]^2 [M^2-k^2]}=\Big\{G_0(w), v_{\mu} G_1(w), g_{\mu \nu} G_2(w)+v_{\mu} v_{\nu} G_3(w)\Big\} ,
\end{equation}
\begin{equation}
	\frac{1}{i} \bigintssss \frac{d^n k}{(2 \pi)^n}  \frac{1}{[v \cdot k-w)[M^2-k^2][(k+q)^2-M^2]} =
	~\bigintssss_0^1 dx \, \frac{1}{2 \widetilde{M}} \, \frac{\partial}{\partial \widetilde{M}} \, J_0 \Big(\widetilde{w}, \widetilde{M}\Big) ,
\end{equation}
\begin{equation}
\begin{aligned}
	   \frac{1}{i} \bigintssss \frac{d^n k}{(2 \pi)^n}  \frac{k_{\mu}}{[v \cdot k-w][M^2-k^2][(k+q)^2-M^2]}  =
	   \bigintssss_0^1 dx \, \Bigg(\frac{ v_{\mu}}{2 \widetilde{M}} \, \frac{\partial}{\partial \widetilde{M}} \, J_1 \Big(\widetilde{w}, \widetilde{M}\Big) 
	                                   - \frac{x \, q_{\mu}}{2 \widetilde{M}} \, \frac{\partial}{\partial \widetilde{M}} \, J_0 \Big(\widetilde{w}, \widetilde{M}\Big)\Bigg) ,
\end{aligned}
\end{equation}
\begin{equation}
	\begin{aligned}
	 & \frac{1}{i} \bigintssss \frac{d^n k}{(2 \pi)^n} \frac{k_{\mu} k_{ \nu}}{[v \cdot k-w][M^2-k^2][(k+q)^2-M^2]} = \\
						  & \bigintssss_0^1 dx \Bigg(\frac{ g_{\mu \nu}}{2 \widetilde{M}} \, \frac{\partial}{\partial \widetilde{M}} \, J_2 \Big(\widetilde{w}, \widetilde{M}\Big) + \frac{v_{\mu} v_{\nu}}{2 \widetilde{M}} \, \frac{\partial}{\partial \widetilde{M}}\, J_3 \Big(\widetilde{w}, \widetilde{M}\Big) 
						- \left(  q_{\mu} v_{\nu} + q_{\nu} v_{\mu} \right)  \frac{x}{2 \widetilde{M}} \, \frac{\partial}{\partial \widetilde{M}} \, J_1 \Big(\widetilde{w}, \widetilde{M}\Big) \\
						  & \qquad \qquad + \frac{x^2 \, q_{\mu} q_{\nu}}{2 \widetilde{M}} \frac{\partial}{\partial \widetilde{M}} \,J_0 \Big(\widetilde{w}, \widetilde{M}\Big) \Bigg) ,  
   \end{aligned}
\end{equation}
where $\widetilde{w}(x)= w + x v \cdot q$, and $\widetilde{M}^2(x)=x(x-1)q^2+M^2$. The analytical expressions for the loop functions in dimensional regularization are   
\begin{equation}
	J_0(w)=~-4Lw+\frac{w}{8 \pi^2}\Bigg[1-2 \, \text{ln}\Bigg(\frac{M}{\lambda}\Bigg)\Bigg]-\frac{1}{4 \pi^2} \sqrt{M^2-w^2} \, \text{ArcCos}\Big(\frac{-w}{M}\Big)+\mathcal{O}(n-4) ,
\end{equation}
for $M^2>w^2$, 
\begin{equation}
J_0(w)=~-4Lw+\frac{w}{8 \pi^2}\Bigg[1-2 \, \text{ln}\Bigg(\frac{M}{\lambda}\Bigg)\Bigg]+\frac{1}{4 \pi^2} \sqrt{w^2-M^2} \, \text{ln}\Bigg(\frac{-w}{M} + \sqrt{ \frac{w^2}{M^2} -1} \Bigg)+\mathcal{O}(n-4) ,
\end{equation}
for $w^2>M^2$, and 
\begin{equation}
J_1(w)=~ wJ_0(w)+\Delta_M, \quad J_2(w)=\frac{1}{n-1} \Bigg[(M^2-w^2) J_0(w)-w\Delta_M \Bigg] ,
\end{equation}
\begin{equation}
J_3(w)=~ wJ_1(w)-J_2(w) ,
\end{equation}
\begin{equation}
G_i(w)=~ \frac{\partial}{\partial w} J_i(w), \quad i=0,1,2,3.
\end{equation}
\end{appendix}

\bibliographystyle{utphysmod}
{\small
\bibliography{bibliography_LambdaB}
}
\end{document}